\documentclass[a4paper,11pt]{article}
\pdfoutput=1 % if your are submitting a pdflatex (i.e. if you have
\usepackage{jcappub} % for details on the use of the package, please
\usepackage[T1]{fontenc} % if needed
\usepackage{enumitem}
\usepackage{verbatim}
\usepackage[T1]{fontenc}
\usepackage[utf8]{inputenc}
\usepackage[american]{babel}
\usepackage{epsfig}
\usepackage{graphicx,subcaption,caption}
\usepackage{booktabs}
\usepackage{multirow}
\usepackage{dcolumn}
\usepackage{amsmath}
\usepackage{mathtools}
\usepackage{amsfonts}
\usepackage{amssymb}
\usepackage{ulem}
\usepackage{epstopdf}
\usepackage{bm}
\usepackage{siunitx}
\usepackage{braket}
\usepackage{enumitem}
\usepackage{soul}

\title{Tracking the Multifield Dynamics with Cosmological Data: A Monte Carlo approach}

\author[a,1]{William Giar\`e, \note{Corresponding author.}}
\author[a]{Mariaveronica De Angelis,}
\author[a]{Carsten van de Bruck,}
\author[a]{Eleonora Di Valentino}

\affiliation[a]{Consortium for Fundamental Physics, School of Mathematics and Statistics, University of Sheffield, Hounsfield Road, Sheffield S3 7RH, United Kingdom}

\emailAdd{w.giare@sheffield.ac.uk}
\emailAdd{mdeangelis1@sheffield.ac.uk}
\emailAdd{c.vandebruck@sheffield.ac.uk}
\emailAdd{e.divalentino@sheffield.ac.uk}

\abstract{We introduce a numerical method specifically designed for investigating generic multifield models of inflation where a number of scalar fields $\phi^K$ are minimally coupled to gravity and live in a field space with a non-trivial metric $\mathcal{G}_{IJ}(\phi^K)$. Our algorithm consists of three main parts. Firstly, we solve the field equations through the entire inflationary period, deriving predictions for observable quantities such as the spectrum of scalar perturbations, primordial gravitational waves, and isocurvature modes. We also incorporate the transfer matrix formalism to track the behavior of adiabatic and isocurvature modes on super-horizon scales and the transfer of entropy to scalar modes after the horizon crossing. Secondly, we interface our algorithm with Boltzmann integrator codes to compute the subsequent full cosmology, including the cosmic microwave background anisotropies and polarization angular power spectra.  Finally, we develop a novel sampling algorithm able to efficiently explore a large volume of the parameter space and identify a sub-region where theoretical predictions agree with observations. In this way, sampling over the initial conditions of the fields and the free parameters of the models, we enable Monte Carlo analysis of multifield scenarios. We test all the features of our approach by analyzing a specific model and deriving constraints on its free parameters. Our methodology provides a robust framework for studying multifield inflation, opening new avenues for future research in the field.}

\begin{document}
\maketitle
\flushbottom

\section{Introduction}
\label{sec.intro}

Inflation is a period of rapid expansion in the early Universe initially proposed to account for various observational phenomena, including spatial flatness, the horizon and entropy problems, and the apparent lack of topological defects~\cite{Guth:1980zm,Linde:1981mu,Albrecht:1982wi}. However, it became quickly clear that inflation also offers an elegant mechanism to explain the physical origins of the first fluctuations in the Universe, which eventually gave rise to the observed structures such as galaxies and clusters of galaxies \cite{Mukhanov:1981xt,Bardeen:1983qw,Hawking:1982cz,Guth:1982ec}. According to the theory of inflation, the initial seeds are generated by quantum fluctuations in the inflaton field (or fields, if inflation is driven by multiple fields) that are stretched to superhorizon scales during inflation, eventually sourcing density fluctuations in other matter species. 

Due to its ability to account for the origin of the structures observed in the present-day Universe, inflation is widely accepted as the leading theory of the very early Universe. However, despite its remarkable success, inflation is not free from limitations. From an observational perspective, the absence of a definitive detection of B-mode polarization~\cite{Planck:2018jri,BICEP:2021xfz} and the emerging discrepancies among different Cosmic Microwave Background (CMB) experiments~\cite{Lin:2019zdn,Forconi:2021que, Handley:2020hdp,LaPosta:2022llv,DiValentino:2022rdg,DiValentino:2022oon,Giare:2022rvg,Calderon:2023obf,Giare:2023xoc} pose new challenges in determining precise predictions for the inflationary models/mechanisms that best explain observational data~\cite{Giare:2023wzl}, and various studies suggest that modifications to the inflationary sector of the cosmological model might play a relevant role in addressing (part of) the tension between the value of the expansion rate of the Universe $H_0$~\cite{Riess:2021jrx,Verde:2019ivm,DiValentino:2020zio,DiValentino:2021izs,Abdalla:2022yfr} measured through direct local distance ladder measurements and the value inferred from CMB observations, see \textit{e.g.},~\cite{DiValentino:2018zjj,Ye:2022efx,Jiang:2022uyg,Jiang:2022qlj,Takahashi:2021bti,Lin:2022gbl,Hazra:2022rdl,Braglia:2021sun,Keeley:2020rmo,Jiang:2023bsz}. On the other hand, from a theoretical standpoint, despite a plethora of proposed models and mechanisms~\cite{Martin:2013nzq}, the nature of the inflation field (or fields) remains still unknown and embedding inflation in a more fundamental theory remains an open problem \cite{Lyth:1998xn,Linde:2007fr,Baumann:2014nda}.

The simplest dynamical models of inflation involve a single scalar field minimally coupled to gravity whose evolution should be governed by a potential enough flat to induce a phase of slow-roll evolution. However, several non-standard realizations of inflation have been proposed both in the context of extensions to the Standard Model of particle physics and in modified gravity theories and tested against a wide range of available data, including CMB, Big Bang Nucleosynthesis, and Gravitational Wave measurement.\footnote{See,\textit{e.g.}, Refs.~\cite{Leach:2002ar,Boubekeur:2005zm,Martin:2006rs,Moss:2007qd,Bezrukov:2010jz,Zhao:2011zb,Martin:2013nzq,Martin:2014rqa,Martin:2014lra,Carrillo-Gonzalez:2014tia,Creminelli:2014oaa,DiValentino:2016nni,DiValentino:2016ziq,Campista:2017ovq,Giare:2019snj,Forconi:2021que,Dai:2019ejv,Baumann:2015xxa,Odintsov:2020ilr,Giare:2020plo,Oikonomou:2021kql,Odintsov:2022cbm,Namba:2015gja,Peloso:2016gqs,Pi:2019ihn,Ozsoy:2020ccy,Stewart:2007fu, Mukohyama:2014gba,Giovannini:2015kfa,Giovannini:2018dob,Giovannini:2018nkt,Giovannini:2018zbf,Giare:2020vhn,Giare:2020vss,Giare:2020plo,Giare:2022wxq,Baumgart:2021ptt,Franciolini:2018ebs,DEramo:2019tit,Giare:2019snj,Caldwell:2018giq,Clarke:2020bil,Caprini_2018,Giare:2020vhn,Allen:1997ad,Smith:2006nka,Boyle:2007zx,Kuroyanagi:2014nba,Ben-Dayan:2019gll,Aich:2019obd,Cabass:2015jwe,Vagnozzi:2020gtf,Benetti:2021uea,Calcagni:2020tvw,Oikonomou:2022ijs,Barrow:1993ad,Peng:2021zon,Ota:2022hvh,Odintsov:2022sdk,Baumgart:2021ptt,Capurri:2020qgz,Canas-Herrera:2021sjs,Odintsov:2023aaw,Oikonomou:2023bah,Fronimos:2023tim,Fronimos:2023tim,Cai:2022lec,Oikonomou:2022irx,Gangopadhyay:2022vgh,Odintsov:2022hxu,Odintsov:2022cbm,Odintsov:2020mkz,Galloni:2022mok,DeAngelis:2021afq,Kallosh:2022ggf,Braglia:2022phb,Vagnozzi:2022qmc} and references therein.} Although a large portion of inflation models or theories is predominantly shaped by a single scalar field, low-energy effective field theories inspired by theories of particle physics beyond the Standard Model or quantum gravity, often incorporate multiple scalar degrees of freedom and suggest that inflation could be driven by multiple fields~\cite{Senatore:2010wk}, potentially featuring non-minimal couplings~\cite{Starobinsky:2001xq,Tsujikawa:2002qx,DiMarco:2002eb,Kaiser:2010ps,Achucarro:2010da,vandeBruck:2010yw,Kaiser:2013sna,vandeBruck:2015tna,vandeBruck:2015xpa,vandeBruck:2016vlw,Carrilho:2018ffi,Achucarro:2019pux,Pinol:2020kvw,Achucarro:2018vey,vandeBruck:2021xkm,DeAngelis:2023fdu}. 
When inflation is driven by multiple scalar fields, after the inflationary period, they have to decay into various standard model particles such as dark matter, baryons, neutrinos, and other species. Furthermore, in multifield models, both adiabatic perturbations and isocurvature modes play crucial roles during inflation~\cite{Weinberg:2004kf}. These modes can persist immediately after the end of inflation and have a significant impact on the evolution of perturbations during the radiation-dominated epoch, giving rise to a rich phenomenology that can be tested and constrained with cosmological and astrophysical data. Therefore, it is important to extract clues about the period of inflation from cosmological observations to determine at least some of the properties of the field(s) driving inflation in the very early Universe~\cite{Tsujikawa:2000wc,Weinberg:2004kf,Kaiser:2010yu,Frazer:2013zoa,Achucarro:2012fd,vandeBruck:2014ata,Dias:2015rca,Dias:2016rjq,Braglia:2020fms,Braglia:2021ckn,Cabass:2022ymb,Geller:2022nkr,Wang:2022eop,Iacconi:2023slv,DeAngelis:2023fdu,Qin:2023lgo,Freytsis:2022aho,Geller:2022nkr,Cicoli:2021yhb,Pinol:2020kvw,Guerrero:2020lng,Garcia-Saenz:2019njm,Nguyen:2019kbm,Li:2019zbk,Bernardeau:2002jy,Kaiser:2012ak,McAllister:2012am,Peterson:2011yt,Dias:2012nf,Kehagias:2012td,Leung:2012ve,Meyers:2010rg,Price:2014ufa,Battefeld:2011yj,Kaiser:2015usz,Ashcroft:2002ap,Paliathanasis:2021fxi,Paliathanasis:2020abu,Paliathanasis:2020wjl,Christodoulidis:2021vye,Piao:2006nm,Rinaldi:2023mdf,Ijaz:2023cvc}. In this regard, it is worth noting that both the amount of isocurvature and adiabatic modes can be accurately constrained by recent measurements of the anisotropies and polarization in the cosmic microwave background radiation~\cite{Planck:2018vyg,Planck:2018nkj,ACT:2020gnv,ACT:2020frw,SPT-3G:2022hvq,SPT-3G:2014dbx,SPT-3G:2021eoc}, offering a valuable opportunity for experimental validation of multifield inflation. However, obtaining precise predictions from generic multifield models/theories is not always easy because observational quantities often depend on various factors. For instance, it is widely known that different initial conditions for the fields lead to different trajectories in field space which can produce slightly different results for observables such as the amplitude of the scalar and tensor perturbations and the spectral index for scalar perturbations, or isocurvature modes~\cite{Easther:2013rva}. This makes a comparison between theory and observations more challenging than in single-field inflation, and most tools employed in cosmological data analyses, such as typical Boltzmann integrator codes and samplers, are either unaware of the physics of inflation or assume single-field potentials.\footnote{A few numerical tools for multifield inflation have been developed, as well. See, \textit{e.g.}, Ref.~\cite{Price:2014xpa}.} As a result, constraining the multifield landscape in light of current observational data represents an ongoing challenge in the field. 

In this paper, we take a first step to tackle down this difficulty and introduce a numerical method to precisely calculate the predictions resulting from generic multifield models of inflation where fields are minimally coupled to gravity and the field space metric is allowed to be non--trivial. To this end, our method is made of three key components.  
First, we numerically solve the complete field equations throughout the entire inflationary period. Once we have ensured that the fields undergo a phase of slow-roll evolution, we numerically solve the background dynamics by adopting a first-order slow-roll approximation and derive predictions for observable quantities such as the spectrum of scalar perturbations, primordial gravitational waves, and isocurvature modes. We also track the behavior of adiabatic and isocurvature modes on super-horizon scales and the transfer of entropy to scalar modes after crossing the horizon using transfer matrix formalism.
Secondly, we interface our algorithm with well-known Boltzmann integrator codes and use the inflationary predictions as initial conditions to compute the subsequent full cosmology, including the CMB anisotropies and polarization angular power spectra. Finally, we narrow down the viable parameter space of the model and derive constraints on its free parameters by introducing a novel sampling algorithm designed to efficiently explore a large parameter volume and identify regions where predictions agree with observations.

The paper is organized as follows. In \autoref{sec.Multifield}, we give an overview of the method, starting with a description of the theoretical parameterization (\autoref{sec.parameterization}), discussing the numerical integration scheme adopted for tracking the multifield dynamics (\autoref{sec.algorithm}), and introducing the sampling algorithm and its interfacing with Boltzmann integrator codes such as \texttt{CAMB}~\cite{Lewis:1999bs,Howlett:2012mh} or \texttt{CLASS}~\cite{Blas:2011rf} (\autoref{sec.TMC}). In \autoref{sec.CaseStudy}, we apply this method to an example inflationary model, obtaining constraints on the parameters of the theory. Our conclusions are presented in \autoref{sec.Conclusion}.

%=======================================================================
\section{Probing the Multifield Landscape}
\label{sec.Multifield}
In this section, we provide a comprehensive overview of the methodology developed for probing the multifield landscape of inflation. In \autoref{sec.parameterization} we describe the theoretical parameterization adopted for the background equations and the formalism used to track the super-horizon evolution of perturbations. In \autoref{sec.algorithm} we introduce our integration algorithm, explaining the various analyses performed to reconstruct the multifield dynamics throughout the full inflationary epoch, and the methodology used for calculating observables such as the spectra of primordial scalar and tensor perturbations, and the entropy transfer after horizon crossing. Finally, in \autoref{sec.TMC}, we explain how our algorithm can be interfaced with standard Boltzmann integrator codes to calculate the subsequent full cosmology of the model and describe the sampling algorithm we designed to explore the parameter-space of generic multifield models and compare theoretical predictions with observations.

\subsection{Parametrizing the Multifield Dynamics}
\label{sec.parameterization}

We start by describing generic multifield models where a number of scalar fields $\phi^K$ are minimally coupled to gravity and live in a field space with a non-trivial metric $\mathcal{G}_{IJ}(\phi^K)$. This geometry is reflected in the kinetic part of the Lagrangian
\begin{equation}
    \mathcal{L}_{kin} = -\frac{1}{2}\, \mathcal{G}_{IJ} \nabla_{\mu}\phi^I\, \nabla ^{\mu} \phi^J.
\end{equation}
Note that in this this paper we always work in units $8\pi G = 1$. Considering a spatially flat Friedmann-Lema\^itre-Robertson-Walker metric, the inflationary dynamics can be described by the generalized Klein-Gordon equations obtained from the variation of the action with respect to the fields $\phi^K$: %, finding
\begin{equation}
   \frac{1}{\sqrt{-g}}\nabla_{\mu}(\sqrt{-g}\, \mathcal{G}_{IJ}\nabla^{\mu}\phi^J) = \frac{1}{2}(\nabla_{\mu}\phi^L)(\nabla^{\mu}\phi^M)\partial_{I}\, \mathcal{G}_{LM} -V_{,I},
   \label{generalkg}
\end{equation}
where $g$ is the determinant of the metric tensor, $V\equiv V(\phi^K)$, and the notation $,_{K}$ indicates the derivative with respect to the fields. 
On the other hand, the evolution of the scale factor $a(t)$ 
is governed by the Friedmann equations:
\begin{equation}
\begin{aligned}
    H^2&=\frac{1}{3}(-\mathcal{L}_{kin}+V(\phi^K)),\\
    \dot{H}&= \mathcal{L}_{kin}.
    \end{aligned}
\end{equation}

To facilitate the interpretation of the evolution of linear cosmological perturbations, we adopt the formalism of Refs.~\cite{Gordon:2000hv,PhysRevD.67.063512} and perform a rotation in the field space. To do so, we define an orthonormal basis in the field space $\{e^I_n\}$ (with $n=1, 2$)  as
\begin{equation}
e^K_n=\frac{\dot{\phi}^K}{\sqrt{2\mathcal{L}_{kin}}},
\end{equation}
(where $|\dot{\phi}^K|\equiv \sqrt{2\mathcal{L}_{kin}}$ denotes the length of the velocity vector $\dot{\phi}^K$ containing the fields as components), and the decomposition of their perturbations \cite{Langlois:2008mn} as
\begin{equation}
    E^K=E^n e_{n}^K.
\end{equation}
In this way, we can introduce the comoving curvature perturbation \cite{Gordon:2000hv,Langlois_2008} 
\begin{equation}
    \zeta = \frac{H}{\sqrt{2\mathcal{L}_{kin}}}E^1,
    \label{zeta}
\end{equation}
which encodes the adiabatic perturbations, as those along the background trajectory in the basis $e^K_1$. It is also worth noting that the whole dynamics of the background fields is encapsulated only in $e^K_1$ and $\dot{e}^K_1$ as $E^K \equiv E^2e_2^K$ vanishes by construction. In this scenario, the rate of \autoref{zeta} clearly depends on the entropy perturbations resulting from the presence of the orthogonal field to the homogeneous trajectory: 
\begin{equation}
    \dot{\zeta}=\frac{H}{\dot{H}}\frac{k^2}{a^2}\Psi + \frac{H}{\mathcal{L}_{kin}}\left(\mathcal{L}_{,2}\,E^2\right),
    \label{zetadot1}
\end{equation}
where $\mathcal{L}_{,2} \equiv e^K_2\,\mathcal{L}_{,K}$ is the projection of $\mathcal{L}_{,K}$ (i.e., of the derivative of total Lagrangian $\mathcal{L}$ with respect to the fields) on $e^K_2$ (i.e., on the direction of the entropic projection of the field acceleration). Clearly, with the presence of the entropy perturbations, the change of $\dot{\zeta}$ could be significant in addition to the general geometry of the field space. 

As previously said, we aim to precisely reconstruct the whole inflationary dynamics and compute the cosmological observables. Therefore, we start by requesting the slow--roll conditions
\begin{equation}
\begin{aligned}
    \mathcal{G}_{IJ}\dot{\phi}^I\dot{\phi}^J &\ll V(\phi^K), \\
    2 \dot{\phi}_I D_t(\dot{\phi}^I)&\ll H \;\mathcal{G}_{IJ}\dot{\phi}^I\dot{\phi}^J.
    \label{slowrollconditions}
\end{aligned}
\end{equation}
In this regime, we can calculate the primordial scalar spectrum predicted by inflation, evaluated at Hubble radius exit, as
\begin{equation}
    \mathcal{P}_{\zeta}(k)\simeq \frac{H^2}{8\pi^2 c_s \,\epsilon}\biggl|_{k=k_*},
\end{equation}
where $c_s$ is the sound speed of the adiabatic perturbation, and
\begin{equation}
\epsilon \equiv -\dot{H}/H^2
\label{epsilon}
\end{equation}
is the usual slow-roll parameter. Since during the slow-roll phase the expansion rate $H$ is almost constant, we expect the spectrum of primordial perturbations to be almost scale-invariant. Therefore one can expand in terms of the small scale-dependence as  
\begin{equation}
   \mathcal{P}_{\zeta}\,(k/k_*)=  A_s \left( \frac{k}{k_*} \right)^{(n_s -1) + \frac{\alpha_s}{2}\ln (k/k_*)+ \frac{\beta_s}{6}\ln^2(k/k_*)},
   \label{Pzeta}
\end{equation}
where $A_s\equiv \mathcal{P}_{\zeta}(k_*)$ is the amplitude of the scalar spectrum at the pivot scale $k_{\star}$, which we fix to $k_{\star}=0.05$ Mpc$^{-1}$ throughout this work. In the simplest parameterization any residual scale dependence of the spectrum is parameterized solely in terms of the scalar spectral index $n_{\rm s}$ defined as:
\begin{equation}
    n_s\equiv \frac{\rm d \ln  \mathcal{P_{\zeta}}}{\rm d \ln k}\biggl|_{k=k_*}=\frac{\rm d \ln \mathcal{P_{\zeta}}}{H \rm dt} \biggl|_{k=k_*}.
    \label{ns}
\end{equation}
However, in \autoref{Pzeta} we also consider higher-order terms, including the running of the spectral index $\alpha_s$ and its running of running $\beta_s$, defined respectively as:
\begin{equation}
 \begin{aligned}
    \alpha_s &\equiv \frac{\rm d n_s}{\rm d \ln k}\biggl|_{k=k_*}=\frac{\text{d} n_s}{H \rm dt}\biggl|_{k=k_*} ,\\
    \beta_s &\equiv \frac{\rm d \alpha_s}{\rm d \ln k}\biggl|_{k=k_*}=\frac{\text{d}  \alpha_s}{H \rm dt}\biggl|_{k=k_*}.
    \end{aligned}
    \label{alphabeta}
\end{equation}

The very same procedure can be repeated for the power spectrum of primordial tensor modes, eventually achieving another useful quantity to constraint general models of inflation, namely the so-called tensor-to-scalar-ratio
\begin{equation}
    r\equiv A_{T}/A_{s},
\end{equation}
which quantifies the fraction of primordial gravitational waves produced by the super-adiabatic amplifications of zero-point quantum fluctuations.

Turning our attention to super-Horizon scales, as seen from \autoref{zetadot1}, once ${k^2}/{a^2} \rightarrow 0$, we eventually get
\begin{equation}
\begin{aligned}
    \dot{\zeta}&= A(t) H(t) \mathcal{S}(t),\\
    \mathcal{\dot{S}}&= B(t)H(t)\mathcal{S}(t),
    \end{aligned}
\end{equation}
with $A$ and $B$ model-dependent functions of time, whose expressions will be explicated in \autoref{IIIA} for a specific model, and
\begin{equation}
    \mathcal{S}=\frac{H}{\sqrt{2\, \mathcal{L}_{kin}}}\,  E^2 ,
    \label{S}
\end{equation}
a dimensionless gauge--invariant quantity. In order to account for the correlation between curvature and isocurvature modes, as well as to estimate the transfer of entropy from the latter to the former during the time from soon after the Hubble exit to the end of inflation, we make use of the transfer matrix formalism~\cite{bib:wands-2002}
\[
\begin{pmatrix}
\zeta\\
\mathcal{S}
\end{pmatrix}
=
\begin{pmatrix}
1 & \mathcal{T}_{\zeta \mathcal{S}}\\
0 & \mathcal{T}_{\mathcal{S}\mathcal{S}}
\end{pmatrix}
\begin{pmatrix}
\zeta\\
\mathcal{S}
\end{pmatrix}_*,
\]
where the transfer functions
\begin{equation}
    \begin{aligned}
        \mathcal{T}_{\zeta \mathcal{S}}(t_*,t)&=\int^t_{t_*} A(t')H(t')\mathcal{T}_{\mathcal{S} \mathcal{S}}(t_*,t')dt',\\
        \mathcal{T}_{\mathcal{S} \mathcal{S}}(t_*,t')&=\text{exp}\left(\int^{t_{'}}_{t_{*}} B(t'')H(t'')dt''\right),
    \label{transf}
        \end{aligned}
\end{equation}
relate the power spectrum at the end of inflation with the power spectrum at the Hubble exit:
\begin{equation}
    \mathcal{P_{\zeta}}=(1+\mathcal{T}_{\zeta\mathcal{S}}^2) \mathcal{P^*_{\zeta}}\equiv \frac{\mathcal{P^*_{\zeta}}}{\cos^2\Theta},
    \label{pr}
\end{equation}
with $\Theta$ being defined as the transfer angle. 

\subsection{Integration Scheme}
\label{sec.algorithm}

After specifying the initial conditions for the fields $\phi^K$, their velocities, and the metric $\mathcal{G}_{IJ}(\phi^K)$ of the field space, we can integrate the equations of motion, \autoref{generalkg}. The integration process is carried out for a maximum number of e-folds, which is set to $N_{\rm{max}}=10000$. During the integration, we dynamically calculate the slow-roll parameter $\epsilon$ by \autoref{epsilon} and continue until the condition $\epsilon=1$ is satisfied. If this condition is not met within $N_{\rm{max}}$ efolds, the model is rejected as inflation does not end. On the other hand, if the condition is satisfied during integration, the point $\epsilon=1$ in the parameter trajectory is considered as a possible end of inflation.

To confirm that it represents the actual end of inflation, we check whether the fields are still active enough to begin a second stage of expansion. Specifically, we test whether the normalized field values with respect to their initial conditions do not exceed a threshold value.\footnote{For the model studied in \autoref{sec.CaseStudy}, we choose $\psi/\psi_{\rm ini} \le 10^{-3}$ and $\chi/\chi_{\rm ini} \le 10^{-2}$.} If the field values satisfy the condition, the multifield dynamics can be considered effectively complete, and the point $\epsilon=1$ reached during integration is regarded as the actual end of inflation. Instead, if the field values do not meet this condition, the integration continues until they fall below the specified threshold. During this stage, we monitor the value of $\epsilon$ to test whether inflation restarts (i.e., if we get back to $\epsilon<1$). If inflation does not restart, then the original point $\epsilon=1$ is set as the end of inflation. If inflation does restart, the new end of inflation is determined by the joint conditions of $\epsilon=1$ and the field values, which must satisfy the selected threshold\footnote{Note that the check detailed in this paragraph is introduced to account for models that exhibit a double inflation behavior, see e.g., Refs.~\cite{Silk:1986vc,Polarski:1992dq,Roberts:1994ap,Langlois:1999dw,Leach:2000yw,Leach:2001zf,Jain:2007au,Jain:2008dw,Jain:2009pm,Kallosh:2014rga,Ragavendra:2020old}. Despite in the prototype model studied in \autoref{sec.CaseStudy} this event appears to be very challenging to realize (i.e., the end of inflation typically corresponds to the conclusion of the field dynamics and does not restart), our code can successfully handle this event, as explicitly discussed in \hyperref[appendix.Double.infl]{Appendix A}.}.

After identifying the end of inflation, we proceed to calculate the total interval of e-fold $\Delta N$ between the start of integration and the end of inflation. We make sure that this interval is greater than a threshold value, which we set at $\Delta N \ge 100$. This threshold is crucial to ensure that inflation lasts long enough to account for the observed homogeneity and isotropy of the Universe, and to establish the appropriate initial conditions for the subsequent Hot Big Bang Theory evolution. 
When the total number of e-folds between the start and end of integration is less than this threshold value, we perform a diagnostic test aimed at determining whether the smaller number of e-folds is due to the initial conditions being chosen too close to the end of inflation or if the model, for that specific combination of parameters, is unable to sustain slow-roll dynamics for an adequate duration. To conduct this test, starting from the same original initial point of integration with the very same initial conditions, we integrate backwards in time from the remaining missing e-folds. During this additional integration, we check the inflationary fields and parameters for any problematic behavior, such as exponentially divergent trajectories, and verify that the model can indeed smoothly support the desired number of e-folds of expansion. If the model fails to meet these requirements, it is deemed unsuitable for explaining the observations and rejected. Note that this test is primarily included to maintain a conservative approach and save a limited number of configurations where models do not respect the threshold value due to the slight shift caused by the random choice of initial conditions while not showing pathological behavior of the backwards-in-time trajectory. This also reduces the impact on the results from the initial conditions themselves, as argued in \autoref{sec.CaseStudy}.\footnote{In the example studied in \autoref{sec.CaseStudy}, most models are ruled out due to their inability to fit observations for the spectral index and the scalar $\&$ tensor amplitudes. The models ruled out because of their inability to sustain a large enough number of e-folds of expansion are the minority, and this is due to the fact that we test from the onset that the fields are settled, so we can safely use the slow-roll approximation when solving the background dynamics.}

After ensuring that the model supports a satisfactory number of e-folds of expansion, and having carefully reconstructed the field dynamics during the entire inflationary phase, we can accurately calculate the entire evolutionary history. This includes how the slow-roll parameters and observables evolve as a function of $N$. By doing so, we can obtain the value of all the slow-roll parameters at horizon crossing, which we choose to be $N_{\star}=55$ e-folds before the end of inflation. Additionally, we take into account the evolution on super-Horizon scales and the transfer of entropy between isocurvature and scalar perturbations. In this way, we can accurately calculate the spectrum of primordial scalar modes (and all the relative observables) at the end of inflation, by means of the transfer matrix formalism detailed in \autoref{sec.parameterization}. 

\subsection{Sampling Method}
\label{sec.TMC}

Once the integration process successfully ends, we can access all the observable predictions of the model, including the amplitude of the scalar perturbation spectrum ($A_s$), its spectral index ($n_s$), the scalar running ($\alpha_{s}$), the scalar running of running ($\beta_{s}$), and the amplitude of the tensor perturbations ($r$). To calculate the subsequent cosmology, we interface our algorithm with standard Boltzmann integrator codes. Specifically, we use the "Code for Anisotropies in the Microwave Background" \texttt{CAMB}~\cite{Lewis:1999bs,Howlett:2012mh}.\footnote{Note that the very same procedure can be used to interface our algorithm with the 'Cosmic Linear Anisotropy Solving System code', \texttt{CLASS}~\cite{Blas:2011rf}.} We input the observable predictions of the multifield inflation as initial conditions for \texttt{CAMB}, along with any other relevant cosmological parameters, such as the other standard $\Lambda$CDM parameters: $\Omega_{\rm b}h^2$, $\Omega_{\rm c}h^2$, $\theta_{\rm{MC}}$, and $\tau$. This allows us to relate the predictions of the multifield model to the usual observable quantities such as the angular power spectra of cosmic microwave background anisotropies and polarization and the matter power spectrum, in both standard and non-standard cosmological backgrounds.

The next crucial step is to explore the parameter space of generic multifield models by sampling over the initial conditions and the free parameters using Monte Carlo techniques. This enables us to compare the theoretical predictions with observational data and derive observational constraints. To accomplish this goal, we have developed a novel sampling algorithm able to explore a sufficiently large volume of the parameter space and identify a sub-region where the model's predictions agree with observations. Our sampling algorithm works as follows. We input a large number of Monte Carlo steps ($\sim 10^6$ steps). At each step, we randomly select the values of the initial conditions and model parameters within some prior ranges for all sampled parameters. Subsequently, we integrate the model for these initial values, performing all the consistency tests detailed in the previous subsection and keeping only models in which the end of inflation is clearly identified and inflation lasts for a sufficiently long number of e-folds able to explain homogeneity and isotropy. If the model satisfies these initial prerequisites, we calculate the observable predictions for all the inflationary parameters, such as  the predicted values of $A_s$, $n_s$, $\alpha_s$, $\beta_s$, and $r$. We then test whether these values fall in reasonable ranges, retaining only models that simultaneously satisfies the following conditions:
\begin{itemize}
\item $A_s \in [1.5\,,\,2.5]\times 10^{-9}$
\item $n_s \in [0.94\, , \, 0.99]$
\item $\alpha_s \in [-0.2\, , \, 0.2]$
\item $\beta_s \in [-0.2\, , \, 0.2]$
\item $r < 0.1$
\end{itemize}
where we have conservatively chosen the previous ranges around the values measured by the most recent cosmic microwave background experiments.\footnote{Note that this step is not mandatory, as we could include all models and assign them a likelihood even if they fall outside these ranges. In practice, models following outside these ranges would have a log likelihood of negative infinity and would not contribute to the observational constraints, resulting in identical results. However, when considering large prior volumes, multifield models can exhibit unpredictable outcomes and we may have several models falling outside this range. Therefore, to prevent the evaluation of the likelihood for models known to be disfavored by data, (which would simply slow down the sampling process without changing anything), we introduce this precautionary measure.} If the model falls within these ranges, we calculate its full cosmology by means of \texttt{CAMB}. In this case, we save as output all the relevant information, including the initial conditions, the values parameters, and all observables. If instead the model fails to satisfy any of these conditions, it is rejected. 

By following this procedure, we generate a chain of models that are equivalent to those produced by typical Markov Chains Monte Carlo (MCMC) techniques. During the sampling process, each saved model is assigned a likelihood based on how well it agrees with the most recent observations of the Cosmic Microwave Background (CMB). Specifically, our reference datasets include:

\begin{itemize}

\item The Planck 2018 temperature and polarization (TT TE EE) data, which also includes low multipole data ($\ell < 30$)~\cite{Aghanim:2019ame,Aghanim:2018eyx,Akrami:2018vks}.

\item The Planck 2018 lensing data, derived from measurements of the power spectrum of the lensing potential~\cite{Aghanim:2018oex}.

\item The latest CMB B-modes power spectrum likelihood cleaned from the foreground contamination as released by Bicep/Keck Array X Collaboration \cite{BICEP:2021xfz}. 

\end{itemize}

To extract a likelihood for each model, starting from these observations we develop an analytical likelihood based on a multi-dimensional normal distribution:
\begin{equation}
\mathcal{L}_{\rm{ike}}\propto 
\rm{exp}\left(-\frac{1}{2} \left(x-\mu\right)^{T} \Sigma^{-1} \left(x-\mu\right) \right)
\label{eq:like}
\end{equation}
where $\mu$ and $\Sigma$ represent the mean values and covariance matrix of parameters obtained by a joint analysis of the aforementioned experiments. We validate our methodology by verifying that both our analytical likelihood and sampler produce consistent results compared to those obtained by using the original experiment likelihoods and publicly available samplers. We refer to \hyperref[appendix.like]{Appendix B} for further details. Consequently, we can obtain informative posterior distributions for the most relevant parameters to be inferred from observations.

%=======================================================================
\section{A Case Study}
\label{sec.CaseStudy}
In this section, we provide a working example of the potentiality of our method. As a case study, we analyze a specific model detailed in \autoref{IIIA}. In \autoref{sec.IIIB} we demonstrate how our algorithm enables us to track the entire fields dynamics and how we can access all the corresponding observables. Finally, in \autoref{sec.IIIC}, we explore the parameter space of the model by sampling over its free parameters and the initial conditions of the fields, deriving observational constraints.

\subsection{Model}
\label{IIIA}

We focus on the simple case where two scalar fields $\phi^K = (\psi,\chi)$ are coupled  through the field metric
\begin{equation}
    \mathcal{G}_{IJ}=\text{diag}\{1, e^{2b(\psi, \chi)}\},
    \label{eq:field_metric}
\end{equation}
and the action reads as
\begin{equation}
     {\cal S} = \int d^4 x \sqrt{-g} \left[ \frac{M_{\rm Pl}^2}{2} R - \frac{1}{2} {\cal G}_{IJ} g^{\mu\nu} \partial_\mu \phi^I \partial_\nu \phi^J - V(\phi^K)  \right],
\end{equation}
where $I, J \in \{1, 2\}$ and $g$ is the determinant of the metric $g_{\mu \nu}$.  These models have been recently studied in Ref.~\cite{DeAngelis:2023fdu} where it was argued that a non-trivial field metric could induce significant changes in the effective mass of the entropy perturbations, opening up a rich phenomenology that we can test for further validations. For the aim of this section, it is worth noting that the fields dynamics is dictated by \autoref{generalkg} which eventually simplifies to~\cite{DeAngelis:2023fdu}
\begin{align}
    \ddot{\psi}+3H\dot{\psi}+V_{,\psi} &=e^{2b(\psi,\chi)}b_{,\psi}\dot{\chi}^2,\label{psikg}\\
    \ddot{\chi}+(3H+2b_{,\psi}\dot{\psi}+b_{,\chi}\dot{\chi})\dot{\chi} &=-e^{-2b(\psi,\chi)}V_{,\chi}.
    \label{chikg}
\end{align}
By performing a rotation in field space, the evolution of the vector along the homogeneous trajectory (the so--called adiabatic field $\sigma$) is given by
\begin{equation}
    \ddot{\sigma}+3H\dot{\sigma}+V_{,\sigma}=0,
\label{sigmakg}
\end{equation}
where $\dot{\sigma}=\sqrt{2\mathcal{L}_{kin}}$. Instead, the entropy part of the equations of motion is given by the rate of change of the angle between the initial field basis ($\phi, \chi$) and the adiabatic/entropy one ($\sigma, s$):
\begin{equation}
     \dot{\theta}=-\frac{V_{,s}}{\dot{\sigma}}-b_{,\psi}\dot{\sigma}\sin \theta.
\label{thetadot}
\end{equation}

To consistently discuss the perturbations of a given cosmological configuration, we take into account the perturbations of the metric $\Phi=\Phi(t,\textbf{x})$ in the longitudinal gauge~\cite{bib:bardeen-1980}
\begin{equation}
    ds^2=-(1+2\Phi)dt^2+a^2(t)(1-2\Phi)d\textbf{x}^2,
    \label{lineelementpert}
\end{equation}
and the corresponding fluctuations of the sources such as $\psi=\psi(t)+\delta \psi_{\textbf{k}}(t,\textbf{x})$ and $ \chi=\chi(t)+\delta \chi_{\textbf{k}}(t,\textbf{x})$. By considering \autoref{lineelementpert} and the perturbed Einstein equations, we write the comoving curvature perturbation \autoref{zeta} in terms of the metric fluctuations as~\cite{DeAngelis:2023fdu}
\begin{equation}
\begin{aligned}
    \zeta&\equiv \Phi-\frac{H}{\dot{H}}(\dot{\Phi}+H\Phi)=\Phi+H\left(\frac{\dot{\psi}\delta\psi+e^{2b}\dot{\chi}\delta\chi}{\dot{\psi}^2+e^{2b}\dot{\chi}^2}\right),
    \label{zeta2}
\end{aligned}
\end{equation}
and its evolution as
\begin{equation}
    \dot{\zeta}= \frac{k^2}{a^2}\frac{H}{\dot{H}}\Phi -2\frac{V_{,s}}{\dot{\sigma}} \mathcal{S},
    \label{zetadot}
\end{equation}
where $\mathcal{S}=\frac{H}{\dot{\sigma}}\delta s$ is the so--called isocurvature perturbation, a gauge-invariant quantity. Note that from \autoref{S} it follows that $E^2 = \delta s$. Clearly, the quantity labelled as adiabatic perturbation \autoref{zetadot} on super-Hubble scales (\emph{i.e.} $k^2/a^2 \rightarrow 0$) is solely fed by the entropy perturbation $\delta s$, namely by the orthogonal field to the background trajectory in field space. Indeed, even if the perturbations in the entropy field evolve independently from the perturbation in the adiabatic field, the large--scale entropy perturbations do impact the evolution of the adiabatic one when the value of the potential curvature is non--zero, i.e. $\eta_{,\sigma s}\neq 0$. Furthermore, their coupling (encoded in the term $V_{,s}$) does not vanish even when a flat field metric is considered. In other words, $\dot{\theta}\equiv 0$. 
For this model, the time--dependent dimensionless functions $A$ and $B$ in the transfer function, \autoref{transf} are~\cite{DeAngelis:2023fdu}
\begin{equation}
\begin{aligned}
    A&=-2\eta_{,\sigma s}+2 \sqrt{2 \epsilon} b_{,\psi}\sin^3 \theta-2\sqrt{2\epsilon}b_{,\chi}e^{-b}\sin^2 \theta \cos \theta,\\
    B&= -\eta_{,ss}+\eta_{,\sigma \sigma}-2\epsilon+\sqrt{2 \epsilon} b_{,\psi}\cos\theta(1+2\sin^2 \theta)+\sqrt{2\epsilon}b_{,\chi} e^{-b} \sin\theta(2\sin^2\theta -1)\\&+\frac{2}{3} \epsilon b_{,\psi}^2+\frac{2}{3} \epsilon b_{,\psi \psi},
    \label{AB}
    \end{aligned}
\end{equation}
where $\eta_{,MN}=V_{,MN}/3H^2$ describes the curvature of the potential in terms of the entropy and adiabatic fields.

These functions encode the coupling between adiabatic and entropy modes and enter in the expressions for the spectral index and its runnings at the end of inflation that, in this general two--field model, can be derived from \autoref{ns}, \autoref{alphabeta}, and \autoref{pr} reading \cite{vandeBruck:2016rfv}
\begin{equation}
\begin{aligned}
n_s &\simeq n_*-2\sin{\Theta}(A_* \cos{\Theta}+B_*\sin{\Theta}),\\
\alpha_s &\simeq\alpha_*+2\cos{\Theta}
 (A_*\cos{\Theta}+B_*\sin{\Theta})\times (A_*\cos{2\Theta}+B_*\sin{2\Theta}),\\
\beta_s&\simeq \beta_*-2\cos{\Theta}(A_*\cos{\Theta}+B_*\sin{\Theta})\times (B_*\cos{2\Theta}-A_*\sin{2\Theta})\\&\times (A_*+2A_*\cos{2\Theta}+B_*\sin{2\Theta}).
\label{index}
\end{aligned}
\end{equation}
From the expressions above we see that constraints on the spectral parameters can be translated into constraints on the geometry of the field metric, namely the curved trajectory in field space between Hubble radius exit and the end of inflation.

\subsection{Predictions}
\label{sec.IIIB}
To be concrete and maintain control over the results, we test all the features of our method by analyzing a simple case where the coupling between the two fields is given by
\begin{equation}
    b(\psi,\chi)=-c \, \frac{\psi \chi}{M^2_{\text{Pl}}},
    \label{eq.coupling}
\end{equation}
with a self-coupling potential of the form
\begin{equation}
    V = \frac{1}{2} m_\psi^2 \psi^2 + \frac{1}{2}m_\chi^2 \chi^2 + g^2\psi^2 \chi^2,
    \label{potential}
\end{equation}
where $m_{\psi}$ and $m_{\chi}$ are the mass terms of the fields and $g$ the coupling constant. 

Once the field metric and the self-coupling potential are specified, the formalism developed in the previous section can be applied and the multifield dynamics numerically solved by means of the integration scheme discussed in \autoref{sec.algorithm}. In particular, the integration of the equations of motion, and therefore the trajectory of the fields, will depend on the model's free parameters (i.e., $m_{\psi}$, $m_{\chi}$, $g$ and $c$) and the initial conditions of the fields (i.e., $\psi_{\rm ini}$, $\dot{\psi}_{\rm ini}$, $\chi_{\rm ini}$ and $\dot{\chi}_{\rm ini}$). In this subsection, we analyze separately the contribution of these parameters to both cosmological observables and the inflationary dynamics, in order to have a comprehensive understanding of their effect and to better interpret the results of the full Monte Carlo analysis performed in the subsequent subsection.

\begin{figure}[t]
\centering
\includegraphics[width=0.75\textwidth]{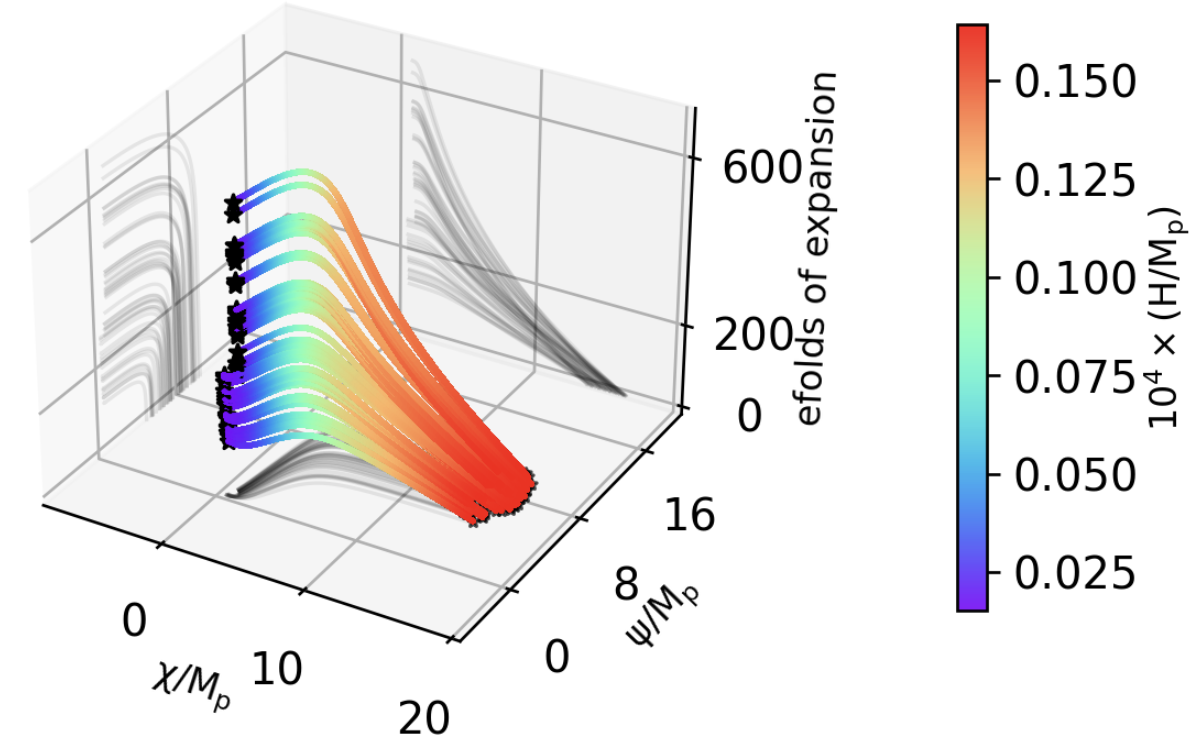}
\caption{\small Field trajectories (as well as  their projections in different 2D-planes in grey) as functions of the e-folds of expansions between the beginning of the integration and the end of inflation. The integration process begins at $N=0$ with randomly selected initial conditions, represented by a black star-like dot in the figure. The end of inflation (marked by another black star-like dot) is determined using the method explained in \autoref{sec.algorithm}. The color-bar shows the value of the Hubble parameter along the field trajectories. For all trajectories, the model's free parameters are fixed to: $m_{\psi}=1.58\times10^{-6}$, $m_{\chi}=3.86\times10^{-6}$, $c=-0.06$, and $g=2\times 10^{-8}$.}
\label{fig:trajectories} 
\end{figure}

We start by examining the robustness of the outcomes of the model when subjected to variations in the initial conditions of the fields. Specifically, we evaluate how the trajectories in the field space change by randomly varying the starting points $\psi_{\text{ini}}$ and $\chi_{\text{ini}}$, while keeping the model's free parameters fixed and setting $\dot{\psi}_{\text{ini}}=\dot{\chi}_{\text{ini}}\simeq 0$. The results are depicted in \autoref{fig:trajectories}. Considering different initial conditions for the fields can result in a period of inflation that is more or less long, while essentially keeping the model's predictions for cosmological observables unchanged. Therefore, we do not expect physical observables to drastically depend on the initial conditions of the fields.

\begin{figure}[t]
	\centering
	\includegraphics[width=0.6\textwidth]{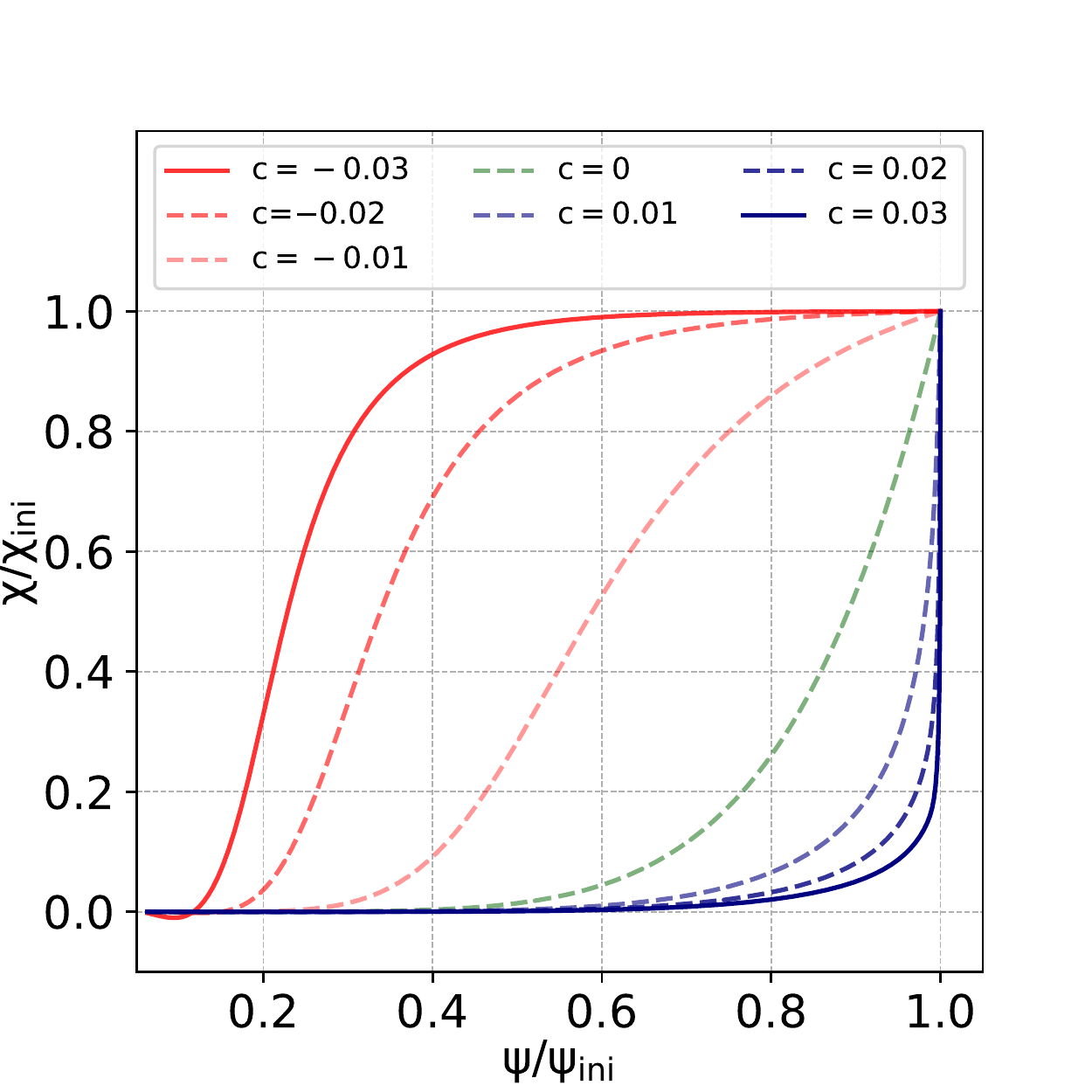}
	\caption{\small Effects of the field metric parameter $c$ on the field evolution in the $\psi$ vs $\chi$ plane. The field values have been normalized to their initial conditions, so the starting point of the trajectories is represented by the coordinates $(1, 1)$, while the coordinates $(0,0)$ ideally represent the end of inflation. The red (blue) curves correspond to negative (positive) values of $c$ as indicated in the legends, while the green curve corresponds to a flat filed metric $\mathcal{G}_{IJ}=\text{diag}\{1,1\},$ (i.e., $c=0$). The other model's free parameters are fixed to: $m_{\psi}=1.58\times10^{-6}$, $m_{\chi}=3.86\times10^{-6}$, %$c=-0.06$, 
    and $g=2\times 10^{-8}$.}
	\label{fig:metric} 
\end{figure}

\begin{figure*}
	\centering
	\includegraphics[width=\textwidth]{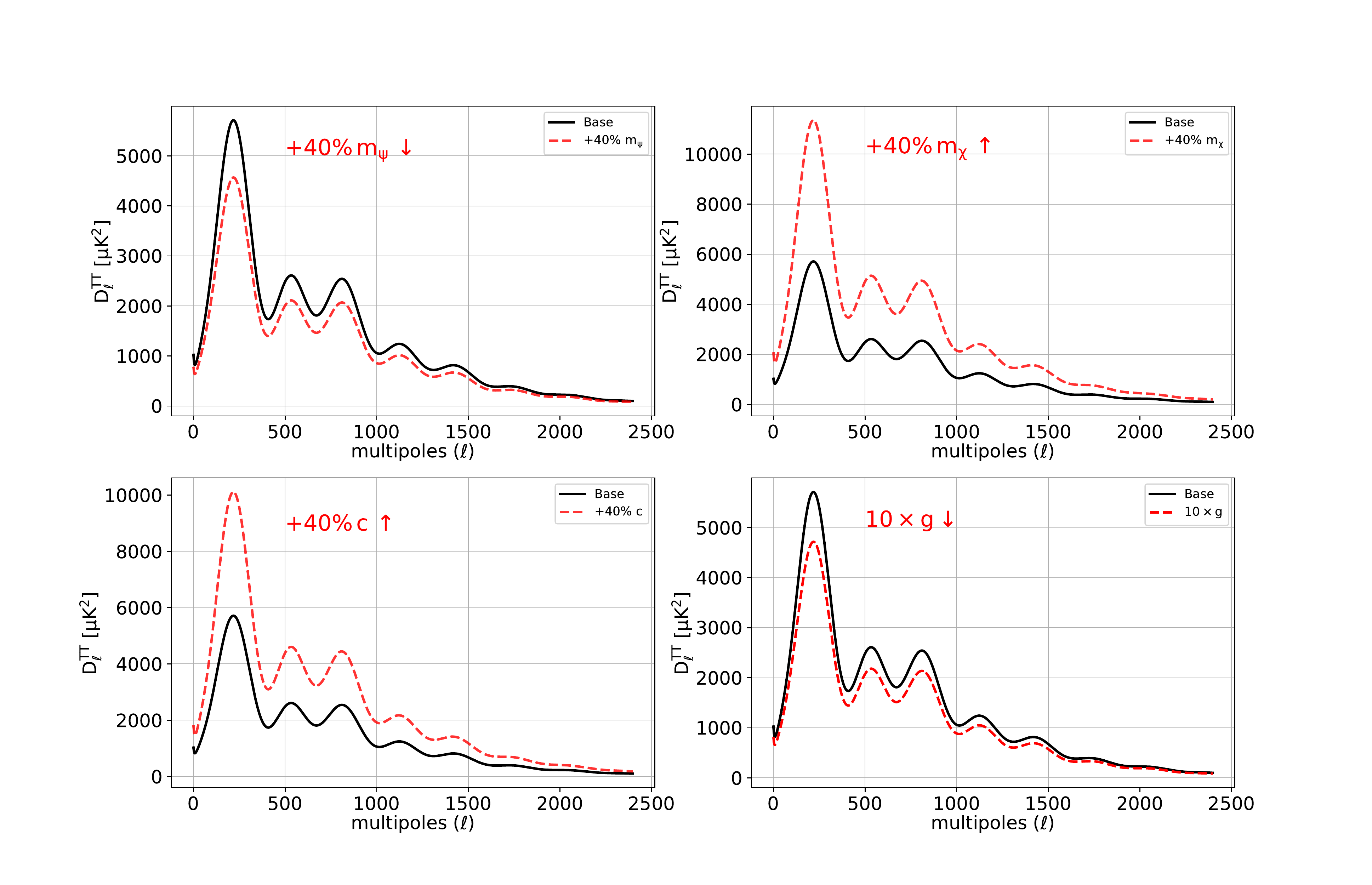}
	\caption{\small Effects on the CMB angular power spectrum resulting from variations in the model's parameters as indicated in the different panels/legends of the figure. The baseline model (black line) corresponds to the following parameter combination: $m_{\psi}=1.58\times10^{-6}$, $m_{\chi}=3.86\times10^{-6}$, $c=-0.06$, and $g=2\times10^{-8}$.}
	\label{fig:spectra} 
\end{figure*}

On the other hand, the variation of the free parameters holds greater significance. For instance, the effects of $c$ which are encoded in the curved field manifold play a substantial role in what concerns the interplay between isocurvature and curvature modes between the horizon crossing and the end of inflation. In particular, when $c$ is gradually negatively reduced, the $\chi$ field velocity remains relatively constant and it becomes stuck for a significant amount of time until it reintegrates into the dynamics when the $\psi$ field reaches the minimum of the potential, see also \autoref{fig:metric}. This effect is translated into the amount of isocurvature mode feeding the curvature one. On the other hand, if $c$ is progressively incremented will lead to a single field scenario as the $\chi$ field drops immediately.\footnote{Interestingly, models that exhibit this peculiar behavior, where a scalar field immediately drops and undergoes a phase of fast oscillations, have been the subject of study as they generate features in primordial density perturbations that directly record the scale factor evolution $a(t)$ while serving as "standard clocks" in the primordial Universe, see e.g., Refs.~\cite{Chen:2012ja,Chen:2014joa,Chen:2014cwa,Chen:2015lza}.}
In other words, the combination of curvature and isocurvature perturbations may lack correlation. However, if the path in field space exhibits curvature from the Hubble exit until the end of inflation, it can introduce an indefinite level of correlation between them impacting the ultimate predictions and potentially causing their growth or reduction, see \autoref{fig:spectra}.

Shifting our focus to the effects of the masses, from \autoref{fig:spectra} we can appreciate how changes in $m_{\chi}$ and $m_{\psi}$ lead to increase or decrease of power in the spectrum of temperatures anisotropies. This can be explained by making explicit \autoref{slowrollconditions} as
\begin{equation}
     V_{,\psi}\simeq-3 \dot{\psi}  H, \qquad V_{,\chi}\simeq-3 \dot{\chi} e^{2b(\psi,\chi)}H,
\end{equation}
and taking into account a self-coupling potential of the form \autoref{potential}, one can easily get
\begin{equation}
    V_{, \chi \chi}=2b_{,\chi}V_{,\chi}-3 e^{2b(\psi,\chi)}\dot{H},
    \label{sr_cond}
\end{equation}
by means of $\frac{dH}{d\chi}= \frac{\dot{H}}{\dot{\chi}}$. Finally, we achieve the following relation  
\begin{equation}
    m_{\chi}^2= 2b_{,\chi}V_{,\chi}+ \frac{3}{2}\dot{\sigma}^2 e^{2b(\psi,\chi)}-2g^2\psi^2.
    \label{m_chicorr}
\end{equation}
Therefore, on the slow--roll trajectory defined by \autoref{sr_cond} there is a correlation between the mass $m_{\chi}$ and the coupling $b_{,\chi}$. 
Such a correlation does not exist for $m_{\psi}$, for which we obtain
\begin{equation}
     m_{\psi}^2= \frac{3}{2}\dot{\sigma}^2 -2g^2\chi^2,
\end{equation}
stating that an increase in $m_{\psi}$ leads to a rapid change in the Hubble rate ($\dot{\sigma}^2 \simeq -\dot{H}$) which translates into a shorter period of inflation and hence a reduction of the amplitude of the power spectrum as seen in \autoref{fig:spectra}. The same figure shows that increasing either $c$ and $m_{\chi}$ results in a larger amplitude of the initial power spectrum, as expected from \autoref{m_chicorr}.

\subsection{Monte Carlo analysis and parameter constraints}
\label{sec.IIIC}

\begin{table*}[ht!]
\begin{center}
\renewcommand{\arraystretch}{1.5}
\begin{tabular}{l@{\hspace{1 cm}}@{\hspace{1 cm}} l @{\hspace{1 cm}} l}
\hline 
\hline 

\textbf{Initial Conditions}  & \textbf{Constraints} & \textbf{Unifrom Prior Ranges} \\
\hline 
$\psi_{\rm ini}/ M_{p}  $                & $-$    & $\psi_{\rm ini}/ M_{p} \in [14\,,\,17]$ \\
$\chi_{\rm ini}/ M_{p}$                & $-$    & $\chi_{\rm ini}/ M_{p} \in [10\,,\,4]$ \\
\hline
\textbf{Model's Parameters}  & \textbf{Constraints} & \textbf{Unifrom Prior Ranges} \\
\hline 
$m_{\psi}$         	           & $< 2.30\cdot 10^{-6}$    &$\log_{10}(m_{\psi}) \in [-8\,,\,-4]$\\
$m_{\chi}$        			   & $<1.01\cdot 10^{-5}$    &$\log_{10}(m_{\chi}) \in [-8\,,\,-4]$\\
$c$         			  			& $< -0.0211$    & $c \in [-1\,,\,1]$\\
$g$             		  			& $<9.72\cdot 10^{-7}$    & $\log_{10}(g) \in [-8\,,\,-5]$\\
\hline 
			
\hline 
\textbf{Primordial spectra}  & \textbf{Constraints} & \\
\hline 
$A_{s}$         & $\left(\,2.109\pm 0.033\,\right)\cdot 10^{-9}$ &\\
$n_s$                        & $0.9621^{+0.0053}_{-0.0047}$ &\\
$\alpha_s$                        & $\left( -0.74^{+0.37}_{-0.32}\right) \times 10^{-3}$&\\
$\beta_s$                        & $\left(-0.103^{+0.088}_{-0.0040}\right) \times 10^{-3}$ &\\
$r$                                    & $<0.04$ & \\

\hline 
\textbf{Entropy Transfer}  & \textbf{Constraints} & \\
\hline 
$\Theta $                      &$< -0.686$    & \\
$A_{\star} $                & $> -1.71$    &  \\
$B_{\star} $                & $> -0.341$    &  \\

\hline
\hline
\end{tabular}
\end{center}
\caption{External priors and observational constraints  at 1$\sigma$ ($68\%$ C.L.) or upper bounds are at 2$\sigma$ ($95\%$ C.L.) on parameters.}
\label{table.Results}
\end{table*}

We use the sampling algorithm detailed in \autoref{sec.TMC} to explore the parameter space of our multifield model. The sampling involves 4 free parameters ($m_{\psi}$, $m_{\chi}$, $g$, and $c$) and the initial conditions of the fields ($\psi_{\rm ini}$, $\chi_{\rm ini}$), which determine the field trajectory during inflation, as pointed out in \autoref{sec.IIIB}. To ensure that we are able to explore a sufficiently large volume of the parameter space, we randomly vary the model parameters and initial conditions in the large uniform priors listed in \autoref{table.Results}. This enables us to obtain a number of combinations of parameters and initial conditions as large as the number of steps in the Monte Carlo analysis. For each step, we use the integration algorithm described in \autoref{sec.algorithm} and compute the full evolution of the fields as well as any observable quantities of the model, including all the primordial scalar and tensor spectrum parameters such as $A_s$, $n_s$, $\alpha_s$, $\beta_s$ and $r$. We test the agreement of each combination of parameter and initial conditions against data by using our likelihood \autoref{eq:like} built on the Planck-2018~\cite{Aghanim:2019ame,Aghanim:2018eyx,Akrami:2018vks,Aghanim:2018oex} and BK18~\cite{BICEP:2021xfz} observations of the cosmic microwave background temperature and polarization anisotropies. 

\begin{figure}[t]
        \centering
	\includegraphics[width=0.75\textwidth]{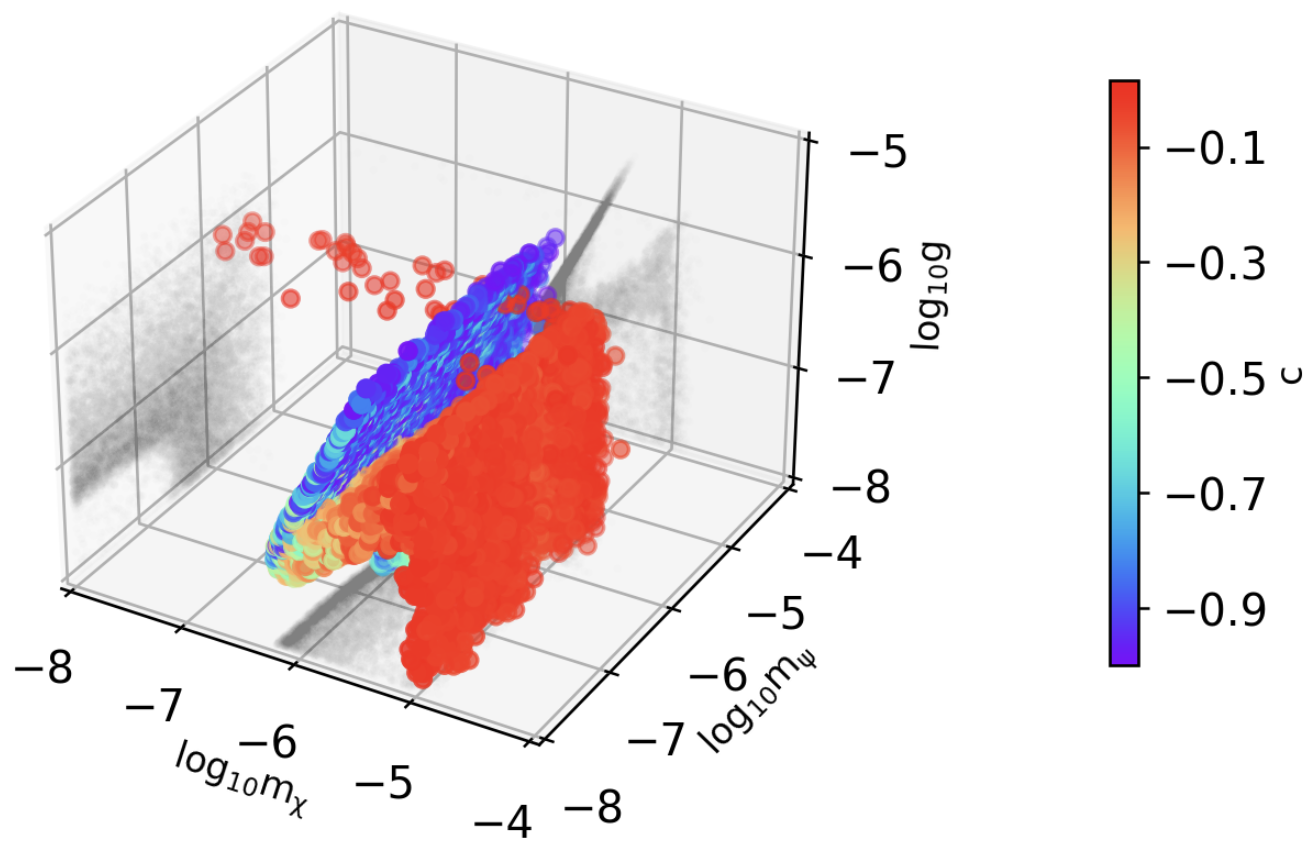}
	\caption{\small How the models distribute in the 4-dimensional parameter space. The box has the size of the prior volume}
	\label{fig:4D_params} 
\end{figure}

\begin{figure*}[t]
	\includegraphics[width=\textwidth]{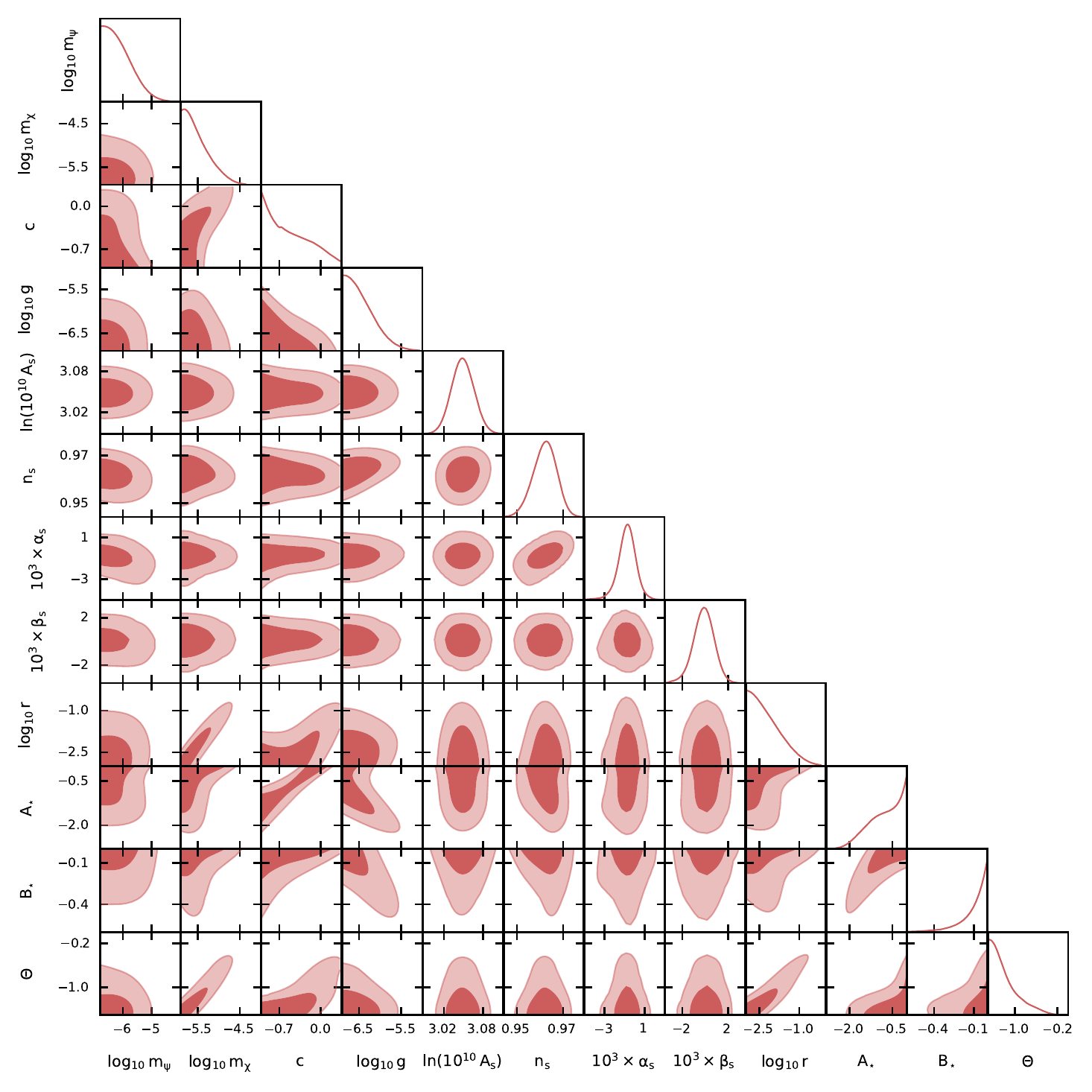}
	\caption{\small Marginalized 2D and 1D posteriors distributions for all the model's parameters and the quantities of interests in this study.}
	\label{fig:Tplot} 
\end{figure*}

Following this methodology, we perform a full Monte Carlo analysis, repeating the process for $\gtrsim 5 \times 10^{6}$ steps and collecting about $2\times 10^4$ sampled models, each of one with its own likelihood. In this way, we derive informative posterior distributions for all the parameters involved in the sampling, as well as for any derived quantities of interest, including those associated with entropy transfer on super-horizon scales. The results are summarized in \autoref{table.Results}, while \autoref{fig:4D_params} depicts the distribution of the sampled models in the 4D parameter space, represented by a box with dimensions corresponding to the size of the prior volume. \autoref{fig:Tplot} shows instead the $68\%$ and $95\%$~CL contour plots for all the quantities of interest in the model.

The first important test we can perform is to study the constraints on the inflationary observables, such as the amplitude of the primordial scalar spectrum $A_s$, the spectral index $n_s$, its runnings $\alpha_s$ and $\beta_s$, and the amplitude of primordial tensor modes $r$. Notice that when dealing with multifield inflation, fixing a model does not necessarily establish consistency relations between these parameters as it usually happens in single-field inflation. Consequently, the specific values of these observables depend on the interplay between the free parameters of the model and the initial conditions of the fields. Additionally, it is important to account for corrections that arise due to the super-horizon evolution of adiabatic and isocurvature perturbations, as discussed in previous sections. Regarding the amplitude of the scalar spectrum, we obtain $A_s=\left(\,2.109\pm 0.033\,\right)\cdot 10^{-9}$ at 68\% CL, in perfect agreement with the model-independent analysis performed with the full Planck and BK18 likelihoods (see also \hyperref[appendix.like]{Appendix B}). Similarly, for the spectral index we get $n_s=0.9621^{+0.0053}_{-0.0047}$ at 68\% CL, while the for the amplitude of primordial gravitational waves we obtain an upped bound $r<0.04$ at 95\% CL. All these results are consistent with the most recent joint analyses of Planck and BK18 data, as well. Regarding the higher-order runnings of the primordial scalar spectrum, we obtain at 68 \% CL $\alpha_s=\left( -0.74^{+0.37}_{-0.32}\right) \times 10^{-3}$ and $\beta_s=\left(-0.103^{+0.088}_{-0.004}\right) \times 10^{-3}$, respectively. Therefore, assuming this particular multifield model, smaller negative values are favored for both the running and the running of the running, although both of them remain consistent with null values at 95\% CL.

One significant aspect of our approach is the ability to obtain observational constraints on the model parameters.  For instance, within this particular model, we are able to place a 95\% CL upper bound on the mass values of the fields that (in Planck units) read $m_{\psi}< 2.30\cdot 10^{-6}$ and $m_{\chi}<1.01\cdot 10^{-5}$, respectively. Similarly, for the coupling parameter $g$ we obtain $g<9.72\cdot 10^{-7}$ at 95\% CL, while for the parameter $c$ controlling the curvature of the filed space we obtain $c< -0.0211$ always at 95\% CL. It is worth noting that these upper bounds assume values significantly far from the limits of the priors adopted for these parameters. In this regard, we have verified that the priors chosen for parameter exploration are sufficiently large to provide uninformative ranges, without introducing any unwanted bias in the parameter constraints. In order to study the correlation between the different parameters, we can refer to \autoref{fig:spectra} for the effects on the angular spectra, \autoref{fig:4D_params} for the correlation in the 4-D parameter space of the model, and \autoref{fig:Tplot} for the contour plot with all sampled and derived parameters. From  \autoref{fig:spectra}, one would expect a correlation between the effects of the parameters $g$ and $c$ on the spectrum of CMB temperature anisotropies as considering more negative values of $c$ leads to a power amplification, while increasing $g$ reduces the amplitude of the spectrum. Therefore, we expect that more negative values of $c$ can be allowed only for larger values of $g$, and this is clearly confirmed by both \autoref{fig:4D_params} and \autoref{fig:Tplot}. Notice also that this model strongly prefers highly negative values of $c$. The reason behind this preference is that when $c$ becomes positive, the curvature of the field space $\mathcal{G}_{IJ}(\phi^K)$ is exponentially suppressed, and one of the two fields essentially becomes a spectator field. As a result, the model reduces to a single-field model with a quadratic potential, which is well-known to predict a significantly large amount of primordial gravitational waves, higher than the current observational limit. Therefore, the parameter space with positive values of $c$ is severely constrained by the B-mode polarization measurements by BK18, in combination with the Planck temperature and polarization measurements.

Finally, our algorithm allows us to derive constraints on any relevant physical quantities in the model, including parameters and functions that govern the transfer of entropy between adiabatic and isocurvature perturbations. For example, we derive a 95\% CL upper bound on the angle $\Theta<-0.686$ appearing in \autoref{index} that weights the corrections acquired by the inflationary parameters between the Hubble exit and the end of inflation. Similarly, we can constrain the time-dependent functions $A(t)$ and $B(t)$ involved in the transfer matrix formalism used to account for the correlation between curvature and isocurvature modes, as well as to estimate the transfer of entropy from the latter to the former during the time from soon after the Hubble exit to the end of inflation. We evaluate these functions at the Hubble exit, finding at 95\%CL the following limits $A_{\star}> -1.71$ and $B_{\star}> -0.341$. Our results confirm that in multi-field models with non-flat metric spaces, the interplay between isocurvature and adiabatic modes plays a crucial role in the final observable predictions, as already argued in Ref.~\cite{DeAngelis:2023fdu}.

%=======================================================================
\section{Conclusion}
\label{sec.Conclusion}
Embedding inflation within a more fundamental framework still remains an open problem, and numerous models and mechanisms have been proposed. The simplest dynamical models of inflation involve a single scalar field minimally coupled to gravity whose evolution is governed by a potential that should be enough flat to induce a phase of slow-roll evolution. However, in certain low-energy effective field theories inspired by string theory, the scalar field sector often comprises multiple scalar fields with non-standard kinetic terms or dynamical couplings. When inflation is driven by multiple scalar fields with non-standard kinetic terms, the interplay between adiabatic perturbations and isocurvature modes becomes significant, influencing the observable predictions and giving rise to a rich phenomenology that can be tested using current cosmological and astrophysical data. In principle, precise measurements of the cosmic microwave background radiation can be used, providing stringent constraints on the abundance of both adiabatic and isocurvature modes and offering a valuable opportunity for experimental validation of these models/theories. However, in practice, obtaining precise predictions from multifield inflation is not always easy as observational quantities often depend on various factors such as the initial conditions of the fields and the specific model assumed. Different trajectories in field space can lead to different results, changing the amplitude of scalar and tensor primordial perturbations. For this reason, most tools employed in cosmological data analyses, such as typical Boltzmann integrator codes and samplers, are either unaware of the physics of inflation or assume single-field potentials. As a result, constraining the multifield landscape with current CMB data represents an ongoing challenge.  

In this work, we take a first step to bridge this gap and introduce an algorithm  specifically designed to investigate generic multifield models of inflation where a number of scalar fields $\phi^K$ are minimally coupled to gravity and live in a field space with a non-trivial metric $\mathcal{G}_{IJ}(\phi^K)$.  In \autoref{sec.Multifield}, we describe both our theoretical parameterization and our numerical method. In particular, after specifying the initial conditions for the fields $\phi^K$, their velocities, the metric $\mathcal{G}_{IJ}(\phi^K)$ of the field space, and the self-coupling potential, our algorithm is able to reconstruct the dynamics of the fields throughout the entire inflationary period precisely determining the end of inflation and performing several consistency tests to ensure the stability of the model. This comprehensive analysis includes more intricate scenarios where distinct field dynamics govern various stages of expansion at different times, such as double inflation or punctuated inflation. An illustrative example can be found in \hyperref[appendix.Double.infl]{Appendix A}. By numerically solving the full field dynamics , we can calculate precise predictions for observable quantities in the slow-roll limit, such as the spectrum of scalar perturbations, primordial gravitational waves, and isocurvature modes. We can also track the super-horizon dynamics of adiabatic and isocurvature modes, determining the transfer of entropy to scalar modes after horizon crossing using the transfer matrix formalism described in \autoref{sec.parameterization}. Once the integration process successfully concludes, we can access all the observable predictions of the model and set the initial conditions to compute the subsequent cosmology. This is done by interfacing our algorithm with standard Boltzmann integrator codes such as \texttt{CAMB} or \texttt{CLASS} that allow us to directly translate the model's predictions in terms of the cosmic microwave background anisotropies and polarization angular power spectra.

Based on this algorithm, we also introduce a novel sampler which is specifically designed to explore a sufficiently large volume of the parameter space of a generic multifield models and identify a sub-region where the model's predictions are in agreement with observations. This allows us to efficiently sample over the initial conditions of the fields and the free parameters of the model, enabling Monte Carlo analysis to compare theoretical predictions with observations. In this work, we make use of the most recent CMB data provided by the Planck collaboration, as well as the latest B-modes power spectrum likelihood released by the Bicep/Keck Array X Collaboration and extract a likelihood for each sampled combinations of parameters. To do so, we develop an analytical likelihood based on these observations that has been extensively tested and proven to reproduce the results obtained from the real likelihoods of different experiments for inflationary parameters, see \hyperref[appendix.like]{Appendix B}.

In \autoref{sec.CaseStudy}, we provide a detailed illustration of our approach by analyzing a specific case study model where two scalar fields are coupled through the field metric by \autoref{eq.coupling}, with a self-coupling potential given by \autoref{potential}. Focusing on this model, in \autoref{sec.IIIB} we use our integration scheme to investigate how the observable predictions change with its parameters and initial conditions. We refer to \autoref{fig:trajectories} and \autoref{fig:metric} for the impact of the field trajectories and the coupling function, respectively. Instead \autoref{fig:spectra} illustrates the impact of the different parameters on the CMB angular power spectra of temperature anisotropies. Finally, in \autoref{sec.IIIC}, we employ our sampler to perform a comprehensive Monte Carlo analysis, deriving observational constraints on the free parameters. The results of our analysis are summarized in \autoref{table.Results}, while \autoref{fig:4D_params} and \autoref{fig:Tplot} show the distribution of sampled models in the parameter space and the correlation among the different parameters, respectively. We are able to derive compelling and precise constraints on both the model's parameters and the inflationary observables such as the primordial power spectra of scalar and tensor perturbations. In addition, we can place constraints on the interplay between curvature and isocurvature modes by accurately accounting for the entropy transfer from isocurvature to curvature perturbations on superhorizon scales. For sake of completeness, in \hyperref[appendix.Double.infl]{Appendix A}, we also discuss the limit where the model reduces to the case of a double quadratic potential with a canonical kinetic term, discussing our ability to recover results already documented in the literature.

Our work provides a robust framework for exploring multifield inflation and opens up exciting opportunities for future research focused on the rich phenomenology expected in both standard and non-standard multifield models of inflation and gravity.

\acknowledgments
\noindent  
CvdB is supported (in part) by the Lancaster–Manchester–Sheffield Consortium for Fundamental Physics under STFC grant: ST/T001038/1. EDV is supported by a Royal Society Dorothy Hodgkin Research Fellowship.
This article is based upon work from COST Action CA21136 Addressing observational tensions in cosmology with systematics and fundamental physics (CosmoVerse) supported by COST (European Cooperation in Science and Technology). We acknowledge IT Services at The University of Sheffield for the provision of services for High Performance Computing. 

\appendix
\section{Double Quadratic Potential $\&$ Double Inflation}
\label{appendix.Double.infl}

In this appendix, we make use of our code to briefly examine a simplified two-field model compared to the one analyzed in \autoref{sec.CaseStudy}. Specifically, we investigate the case of a double-field quadratic potential with a canonical kinetic term 
\begin{equation}
  V = \frac{1}{2} m_\psi^2 \psi^2 + \frac{1}{2}m_\chi^2 \chi^2
\end{equation}
which falls within the parameterization adopted in \autoref{sec.CaseStudy} once we fix $c=0$ and $g=0$. 

Notice that this model has been already discussed in the literature and a comprehensive systematic review of its properties is beyond the goal of our work. That being said, this appendix serves a dual purpose: on one side, we aim to use this simplified case as a safety check to demonstrate our ability to recover many of the results already documented in the literature. On the other hand, we take this opportunity to provide a working example of some features of our algorithm, such as its ability to correctly identify the end of inflation and effectively handle scenarios of double inflation where two stages of expansions are driven by distinct fields at distinct times\footnote{It is worth noting that these scenarios are more commonly realized in models of double quadratic potential with canonical kinetic terms than in the models detailed in \autoref{sec.CaseStudy}.}.

\begin{figure*}[htbp]
\centering
        \includegraphics[width=0.43\textwidth]{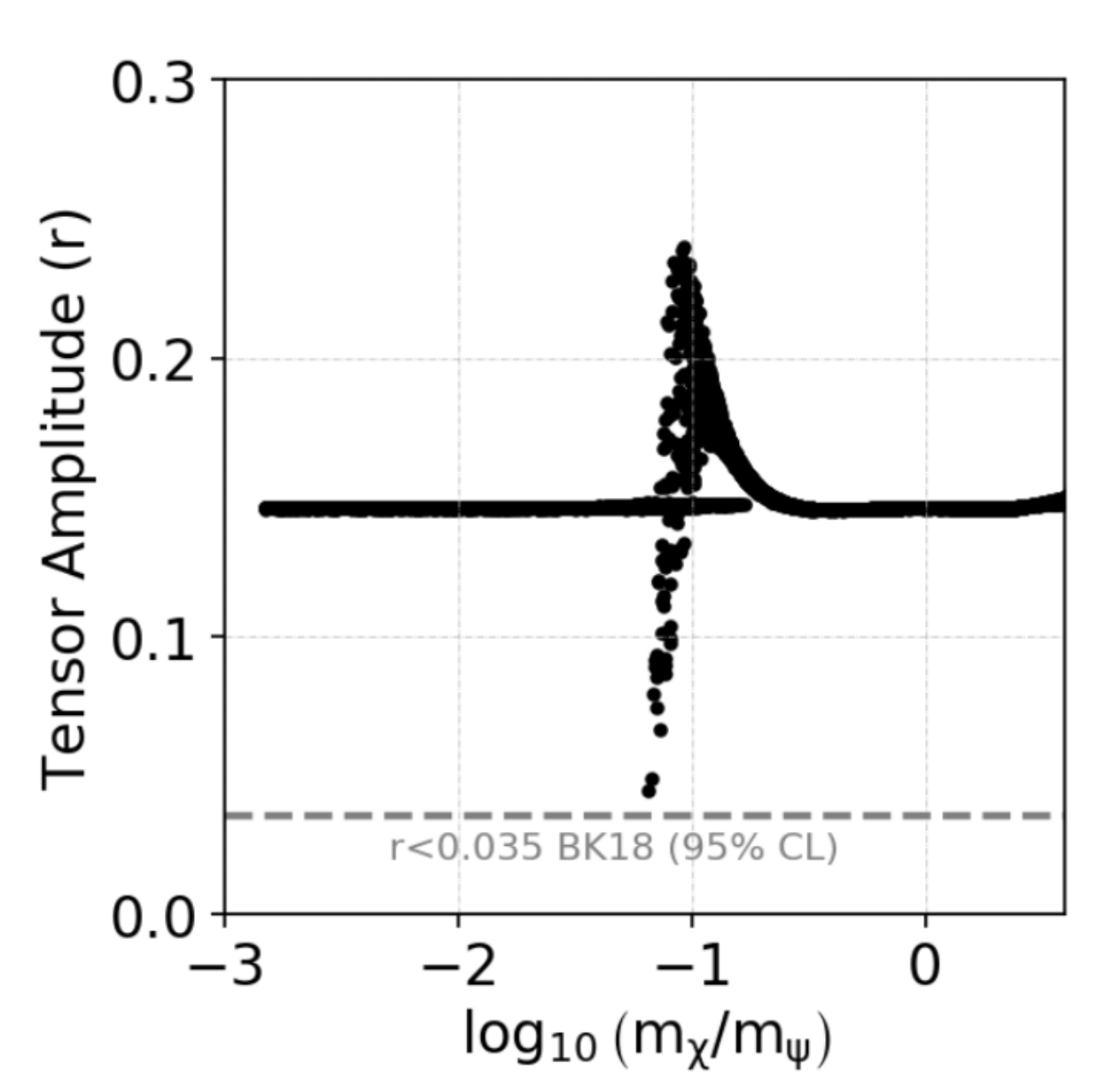}
	\caption{\small Predictions for the tensor amplitude $r$ in terms of the mass ratio $\log_{10}(m_{\chi} / m_{\psi})$ for $\sim 10^4$ models where the amplitude of the scalar spectrum $A_s$ and its tilt $n_s$ fall within ranges consistent with our observational data.}
	\label{fig:double_quadratic_case} 
\end{figure*}

Regarding the topic of interest for our discussion, an analysis similar to the one we carry on here is given in Ref.~\cite{Peterson:2010np} where, using a parameterization for the inflationary dynamics very close to the one we have developed in this paper, the authors point out the characteristics of this two-field model, deriving predictions for observables such as the amplitude of primordial spectra. As highlighted in Section III.C of Ref.~\cite{Peterson:2010np}, a distinctive feature of this potential is the prediction of a tensor-to-scalar-ratio $r\gtrsim 0.13$, regardless of the initial conditions for the fields and the values of the mass ratio. Given that a similar value for the tensor amplitude does not appear to be in agreement with the most recent measurements of CMB B-mode polarization released by the Keck Array collaboration, we do not perform a data analysis (which would be inconclusive given the model's inability to reconcile with the latest observations). However, we perform a model sampling as a cross-validation of the predictions for $r$. As seen in \autoref{fig:double_quadratic_case}, collecting approximately $10^4$ models where the amplitude of the scalar spectrum $A_s$ and its tilt $n_s$ fall within ranges consistent with our observational data, we find that the the tensor amplitude $r$ remains consistently above the $95\%$ limit resulting from the joint Planck+BK18 analysis. Furthermore, the value of $r$ predicted by this potential remains largely unchanged both with respect to the mass ratio (across multiple orders of magnitude) and with respect to the initial conditions (which are randomly chosen from model to model) as well as in excellent agreement with Figure 8 of Ref~\cite{Peterson:2010np}.

A closer analysis of the figure reveals the presence of a minor dispersion of patterns that deviate from the degeneration line between the mass ratio and the value of $r$. A detailed analysis of these models has revealed that this small dispersion is due to double inflation. While theoretically investigated, in Ref.~\cite{Peterson:2010np} these models were not given thorough consideration and the inflationary dynamics were computed by treating double inflation as if it consists of only one inflationary phase. However, as explained in \autoref{sec.algorithm}, our algorithm has been designed to accurately identify all inflationary phases and comprehensively reconstruct their dynamics. As a result,  we can provide precise predictions for scenarios involving double inflation, as well.

\begin{figure*}[htbp]
\centering
        \includegraphics[width=0.5\textwidth]{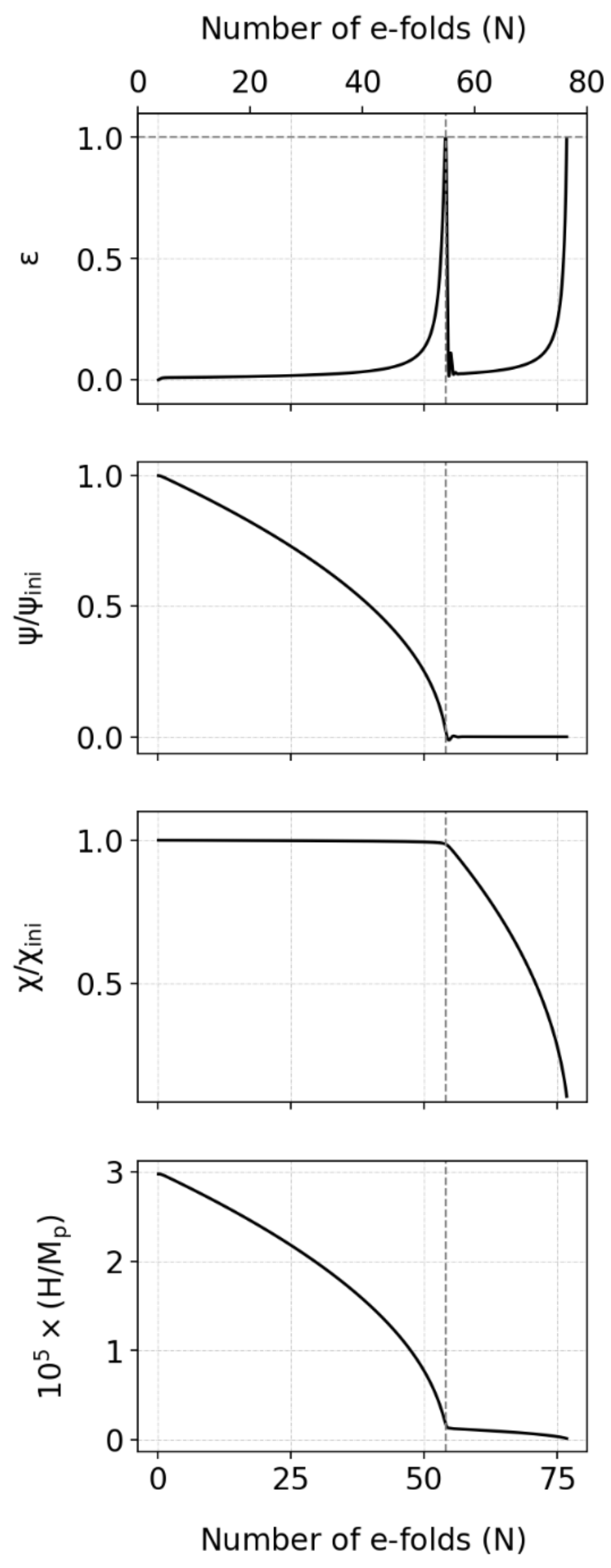}
	\caption{\small 
Evolution of the parameter $\epsilon$ (top panel), the scalar fields $\psi$ and $\chi$ (second and third panel, respectively), and the Hubble parameter $H$ (bottom panel) in a model of double inflation characterized by two stages of expansion.}
	\label{fig:double_inf} 
\end{figure*}

To put it more quantitatively and clarify our treatment of double inflation, we focus on one specific model among those depicted in \autoref{fig:double_quadratic_case}. Namely, we consider the one which shows the smallest tensor amplitude ($r\sim 0.05$, yet outside the Planck-BK18 range) and provide full details of the dynamics of the field and the background evolution in \autoref{fig:double_inf}. As evident from the top panel of the figure, in this model, the parameter $\epsilon$ reaches the critical value $\epsilon=1$ for the first time after about 55 e-folds (grey dashed line). During this initial phase, the expansion of the Universe is driven by the field $\psi$, whose evolution is depicted in the second panel of the same figure. As one can observe, when $\epsilon=1$, the dynamics of $\psi$ is mostly completed, while the second field $\chi$ (whose evolution is shown in the third panel) remains mostly inactive during this phase. Once the inflationary stage guided by $\psi$ terminates, our algorithm monitors the dynamics of the fields and, since $\chi$ is still very active, the integration of the equation of motion continues until the second field also completes its dynamics. This allows us to identify a second inflationary phase, this time guided by $\chi$. In both of these stages, the background dynamics, represented by the evolution of the Hubble parameter $H$ (bottom panel), is accurately computed together with all observational predictions, including the tensor amplitude.

\section{Sampling and Likelihood Validation}
\label{appendix.like}
In this appendix, we provide a step by step explanation of the methodology used to build our likelihood based on the joint analysis of B-Modes polarization data from BK18 and the Planck-2018 measurements of temperature and polarization anisotropies. Most importantly, we prove that our method/likelihood is able to reproduce the same results obtained by the most commonly Markov Chain Monte Carlo (MCMC) analyses performed in the literature. To do this, through all this appendix, we do not consider any explicit inflationary models, and remain completely agnostic about the physics of inflation, as customary in the literature. \\

\begin{table}[t]
	\begin{center}
		\renewcommand{\arraystretch}{1.5}
		\begin{tabular}{l@{\hspace{0. cm}}@{\hspace{1 cm}} c @{\hspace{1 cm}} c}
			\hline
			\textbf{Parameter}    & \textbf{Real likelihoods} & \textbf{This work} \\
			\hline\hline
			$\log(10^{10}A_{\rm S})$     & $3.049\pm 0.016$ & $3.051\pm 0.015$ \\
			$n_{\rm s}$                  & $0.9624\pm 0.0044$ & $0.9621\pm 0.0046$\\
                $\alpha_{\rm s}$             & $0.002\pm 0.010$ & $0.002\pm 0.010$  \\
			$\beta_{\rm s}$              & $0.012\pm 0.012$ & $0.013\pm 0.013$ \\
			$r$                          & $<0.0354$ & $< 0.0357$ \\
			\hline\hline
		\end{tabular}
		\caption{\small Results for the $\Lambda$CDM+$\alpha_s$+$\beta_s$+$r$ model obtained using the publicly available sampler \texttt{Cobaya}~\cite{Torrado:2020xyz} in combination with the full Planck 2018~\cite{Aghanim:2019ame,Aghanim:2018eyx,Akrami:2018vks,Aghanim:2018oex} and BK18~\cite{BICEP:2021xfz} likelihoods (referred to as 'Real likelihoods'), and the results for the same model derived using our sampling algorithm in combination with our analytical likelihood (\autoref{eq:like}) (referred to as 'This work').}
		\label{tab.LCDM_r_nrun_nrunrun}
	\end{center}
\end{table}

\begin{figure*}[htbp]
\centering
	\includegraphics[width=0.7\textwidth]{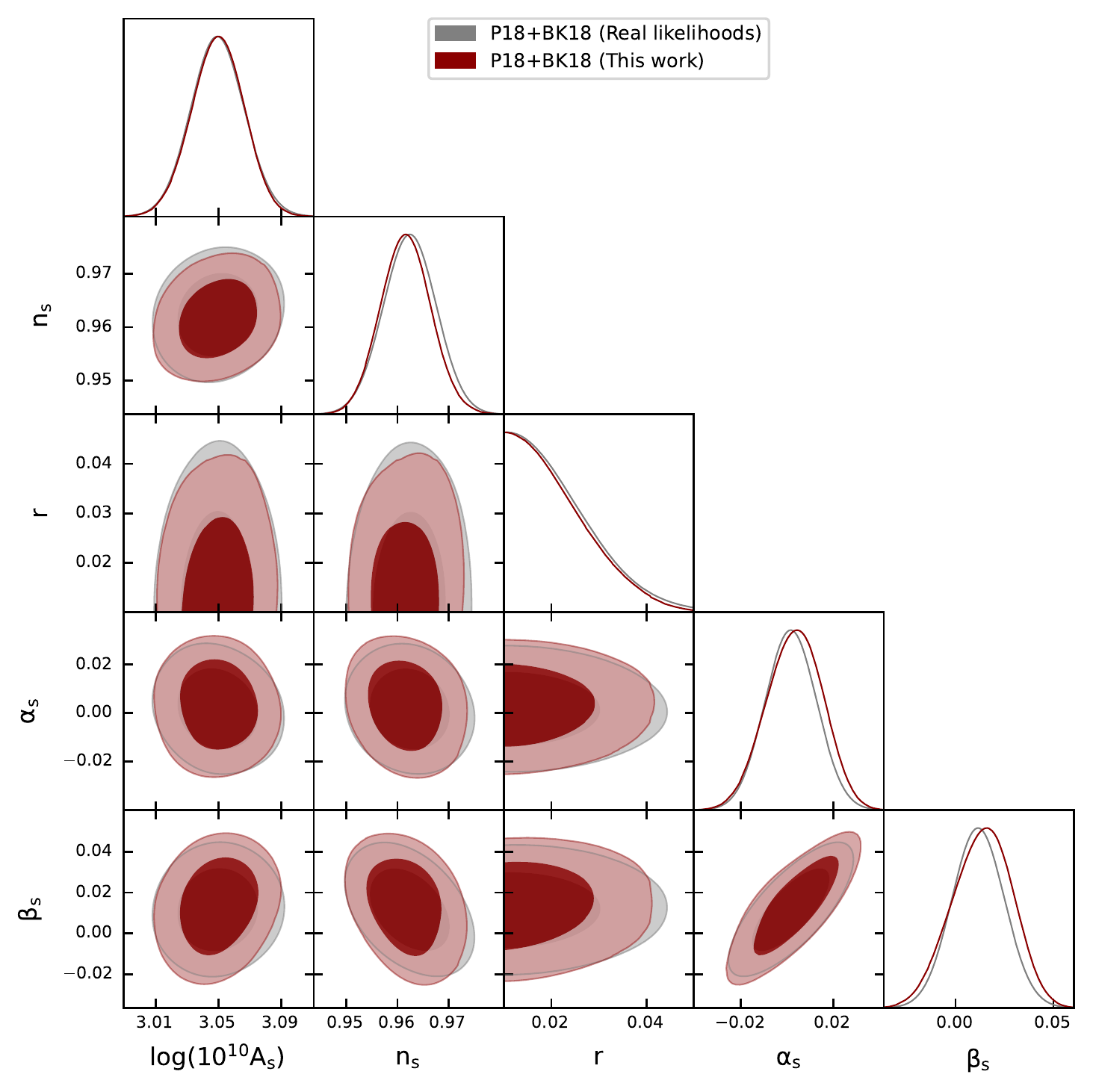}
	\caption{\small  Marginalized 2D and 1D posteriors distributions for the $\Lambda$CDM+$\alpha_s$+$\beta_s$+$r$ model obtained using the publicly available sampler \texttt{Cobaya}~\cite{Torrado:2020xyz} in combination with the full Planck 2018~\cite{Aghanim:2019ame,Aghanim:2018eyx,Akrami:2018vks,Aghanim:2018oex} and BK18~\cite{BICEP:2021xfz} likelihoods (grey), and our sampling algorithm in combination with our analytical likelihood (red). }
	\label{fig:like} 
\end{figure*}

1) As a first step, we consider an extension to the standard cosmological model which includes three additional parameters: the tensor amplitude $r$, the running of the spectral index $\alpha_s$, and the running of the running $\beta_s$. We refer to this model as $\Lambda$CDM+$\alpha_s$+$\beta_s$+$r$ and perform a full Monte Carlo Markov Chain (MCMC) analysis using the publicly available sampler \texttt{Cobaya}~\cite{Torrado:2020xyz}, and Boltzmann integrator code \texttt{CAMB}~\cite{Lewis:1999bs,Howlett:2012mh}. \texttt{Cobaya} explores the posterior distributions of the parameter space using the MCMC sampler developed for \texttt{CosmoMC}\cite{Lewis:2002ah}, which has been specifically adapted for parameter spaces with a hierarchy of speeds by implementing the "fast dragging" procedure introduced in Ref.~\cite{Neal:2005}. The baseline datasets involved in our MCMC analysis consist of the Planck 2018 temperature and polarization (TT TE EE) likelihood, which includes low multipole data ($\ell < 30$)~\cite{Aghanim:2019ame,Aghanim:2018eyx,Akrami:2018vks}, as well as the lensing likelihood obtained from measurements of the power spectrum of the lensing potential\cite{Aghanim:2018oex}. Additionally, we include the most recent CMB B-modes power spectrum likelihood cleaned from foreground contamination, as released by the Bicep/Keck Array X Collaboration \cite{BICEP:2021xfz}. The resulting constraints are presented in \autoref{tab.LCDM_r_nrun_nrunrun}, and we also take this opportunity to update the current bounds on this cosmological extension with the latest data.\\

2) As a second step, using the results from our MCMC analyses, we construct our analytical likelihood based on \autoref{eq:like}. In particular, we adopt the generalized covariance matrix $\Sigma$ and the mean values $\mu$ of parameters obtained for the $\Lambda$CDM+$\alpha_s$+$\beta_s$+$r$ model.\\

3) Finally, we test that our likelihood is able to reproduce the same constraints as real likelihoods from experiments. This is a crucial step needed to validate the results obtained when our likelihood is applied to the study of physical models of multifield inflation. To prove the equivalence of our method and the MCMC analysis, we perform a consistency check for the same $\Lambda$CDM+$\alpha_s$+$\beta_s$+$r$ cosmological model. Specifically, we perform a new run varying the inflationary parameters \{$A_s$, $n_s$, $\alpha_s$, $\beta_s$\} using our sampling algorithm.  In this case, the sampling is performed directly in the parameter space of inflationary observables and using the same prior ranges used for the MCMC analyses. Since sampling on the inflationary parameters requires significantly fewer computational resources, avoiding also some non-physical behaviors that may arise when exploring non-standard inflationary models or solving the inflationary dynamics, we are able to accumulate $\sim 10^5$ models, each of them evaluated using our analytical likelihood \autoref{eq:like}. In \autoref{fig:like}, we directly compare the results obtained using our approach (labeled as 'this work') to the results obtained using the real likelihoods from experiments and the sampler \texttt{Cobaya} (labeled as 'Real likelihoods'). The two methods yield the same constraints for all inflationary parameters, see also \autoref{tab.LCDM_r_nrun_nrunrun}.

%\begin{thebibliography}
\bibliographystyle{unsrt}
\bibliography{bib}

\begin{thebibliography}{100}

\bibitem{Guth:1980zm}
Alan~H. Guth.
\newblock {The Inflationary Universe: A Possible Solution to the Horizon and
  Flatness Problems}.
\newblock {\em Phys. Rev. D}, 23:347--356, 1981.

\bibitem{Linde:1981mu}
Andrei~D. Linde.
\newblock {A New Inflationary Universe Scenario: A Possible Solution of the
  Horizon, Flatness, Homogeneity, Isotropy and Primordial Monopole Problems}.
\newblock {\em Phys. Lett. B}, 108:389--393, 1982.

\bibitem{Albrecht:1982wi}
Andreas Albrecht and Paul~J. Steinhardt.
\newblock {Cosmology for Grand Unified Theories with Radiatively Induced
  Symmetry Breaking}.
\newblock {\em Phys. Rev. Lett.}, 48:1220--1223, 1982.

\bibitem{Mukhanov:1981xt}
Viatcheslav~F. Mukhanov and G.~V. Chibisov.
\newblock {Quantum Fluctuations and a Nonsingular Universe}.
\newblock {\em JETP Lett.}, 33:532--535, 1981.

\bibitem{Bardeen:1983qw}
James~M. Bardeen, Paul~J. Steinhardt, and Michael~S. Turner.
\newblock {Spontaneous Creation of Almost Scale - Free Density Perturbations in
  an Inflationary Universe}.
\newblock {\em Phys. Rev. D}, 28:679, 1983.

\bibitem{Hawking:1982cz}
S.~W. Hawking.
\newblock {The Development of Irregularities in a Single Bubble Inflationary
  Universe}.
\newblock {\em Phys. Lett. B}, 115:295, 1982.

\bibitem{Guth:1982ec}
Alan~H. Guth and S.~Y. Pi.
\newblock {Fluctuations in the New Inflationary Universe}.
\newblock {\em Phys. Rev. Lett.}, 49:1110--1113, 1982.

\bibitem{Planck:2018jri}
Y.~Akrami et~al.
\newblock {Planck 2018 results. X. Constraints on inflation}.
\newblock {\em Astron. Astrophys.}, 641:A10, 2020.

\bibitem{BICEP:2021xfz}
P.~A.~R. Ade et~al.
\newblock {Improved Constraints on Primordial Gravitational Waves using Planck,
  WMAP, and BICEP/Keck Observations through the 2018 Observing Season}.
\newblock {\em Phys. Rev. Lett.}, 127(15):151301, 2021.

\bibitem{Lin:2019zdn}
Weikang Lin and Mustapha Ishak.
\newblock {A Bayesian interpretation of inconsistency measures in cosmology}.
\newblock {\em JCAP}, 05:009, 2021.

\bibitem{Forconi:2021que}
Matteo Forconi, William Giar\`e, Eleonora Di~Valentino, and Alessandro
  Melchiorri.
\newblock {Cosmological constraints on slow roll inflation: An update}.
\newblock {\em Phys. Rev. D}, 104(10):103528, 2021.

\bibitem{Handley:2020hdp}
Will Handley and Pablo Lemos.
\newblock {Quantifying the global parameter tensions between ACT, SPT and
  Planck}.
\newblock {\em Phys. Rev. D}, 103(6):063529, 2021.

\bibitem{LaPosta:2022llv}
Adrien La~Posta, Umberto Natale, Erminia Calabrese, Xavier Garrido, and Thibaut
  Louis.
\newblock {Assessing the consistency between CMB temperature and polarization
  measurements with application to Planck, ACT, and SPT data}.
\newblock {\em Phys. Rev. D}, 107(2):023510, 2023.

\bibitem{DiValentino:2022rdg}
Eleonora Di~Valentino, William Giar\`e, Alessandro Melchiorri, and Joseph Silk.
\newblock {Quantifying the global \textquoteleft{}CMB tension\textquoteright{}
  between the Atacama Cosmology Telescope and the Planck satellite in extended
  models of cosmology}.
\newblock {\em Mon. Not. Roy. Astron. Soc.}, 520(1):210--215, 2023.

\bibitem{DiValentino:2022oon}
Eleonora Di~Valentino, William Giar\`e, Alessandro Melchiorri, and Joseph Silk.
\newblock {Health checkup test of the standard cosmological model in view of
  recent cosmic microwave background anisotropies experiments}.
\newblock {\em Phys. Rev. D}, 106(10):103506, 2022.

\bibitem{Giare:2022rvg}
William Giar\`e, Fabrizio Renzi, Olga Mena, Eleonora Di~Valentino, and
  Alessandro Melchiorri.
\newblock {Is the Harrison-Zel'dovich spectrum coming back? ACT preference for
  $n_s \sim 1$ and its discordance with Planck}.
\newblock 10 2022.

\bibitem{Calderon:2023obf}
Rodrigo Calder\'on, Arman Shafieloo, Dhiraj~Kumar Hazra, and Wuhyun Sohn.
\newblock {On the consistency of $\Lambda$CDM with CMB measurements in light of
  the latest Planck, ACT, and SPT data}.
\newblock 2 2023.

\bibitem{Giare:2023xoc}
William Giar\`e.
\newblock {CMB Anomalies and the Hubble Tension}.
\newblock 5 2023.

\bibitem{Giare:2023wzl}
William Giar\`e, Supriya Pan, Eleonora Di~Valentino, Weiqiang Yang, Jaume
  de~Haro, and Alessandro Melchiorri.
\newblock {Inflationary Potential as seen from Different Angles: Model
  Compatibility from Multiple CMB Missions}.
\newblock 5 2023.

\bibitem{Riess:2021jrx}
Adam~G. Riess et~al.
\newblock {A Comprehensive Measurement of the Local Value of the Hubble
  Constant with 1 km/s/Mpc Uncertainty from the Hubble Space Telescope and the
  SH0ES Team}.
\newblock {\em Astrophys. J. Lett.}, 934(1):L7, 2022.

\bibitem{Verde:2019ivm}
L.~Verde, T.~Treu, and A.~G. Riess.
\newblock {Tensions between the Early and the Late Universe}.
\newblock In {\em {Nature Astronomy 2019}}, 2019.

\bibitem{DiValentino:2020zio}
Eleonora Di~Valentino et~al.
\newblock {Cosmology Intertwined II: The Hubble Constant Tension}.
\newblock {\em arXiv:2008.11284}, 8 2020.

\bibitem{DiValentino:2021izs}
Eleonora Di~Valentino, Olga Mena, Supriya Pan, Luca Visinelli, Weiqiang Yang,
  Alessandro Melchiorri, David~F. Mota, Adam~G. Riess, and Joseph Silk.
\newblock {In the realm of the Hubble tension\textemdash{}a review of
  solutions}.
\newblock {\em Class. Quant. Grav.}, 38(15):153001, 2021.

\bibitem{Abdalla:2022yfr}
Elcio Abdalla et~al.
\newblock {Cosmology intertwined: A review of the particle physics,
  astrophysics, and cosmology associated with the cosmological tensions and
  anomalies}.
\newblock {\em JHEAp}, 34:49--211, 2022.

\bibitem{DiValentino:2018zjj}
Eleonora Di~Valentino, Alessandro Melchiorri, Yabebal Fantaye, and Alan
  Heavens.
\newblock {Bayesian evidence against the Harrison-Zel\textquoteright{}dovich
  spectrum in tensions with cosmological data sets}.
\newblock {\em Phys. Rev. D}, 98(6):063508, 2018.

\bibitem{Ye:2022efx}
Gen Ye, Jun-Qian Jiang, and Yun-Song Piao.
\newblock {Toward inflation with ns=1 in light of the Hubble tension and
  implications for primordial gravitational waves}.
\newblock {\em Phys. Rev. D}, 106(10):103528, 2022.

\bibitem{Jiang:2022uyg}
Jun-Qian Jiang and Yun-Song Piao.
\newblock {Toward early dark energy and ns=1 with Planck, ACT, and SPT
  observations}.
\newblock {\em Phys. Rev. D}, 105(10):103514, 2022.

\bibitem{Jiang:2022qlj}
Jun-Qian Jiang, Gen Ye, and Yun-Song Piao.
\newblock {Return of Harrison-Zeldovich spectrum in light of recent
  cosmological tensions}.
\newblock 10 2022.

\bibitem{Takahashi:2021bti}
Fuminobu Takahashi and Wen Yin.
\newblock {Cosmological implications of ns~1 in light of the Hubble tension}.
\newblock {\em Phys. Lett. B}, 830:137143, 2022.

\bibitem{Lin:2022gbl}
Chia-Min Lin.
\newblock {D-term inflation in braneworld models: Consistency with
  cosmic-string bounds and early-time Hubble tension resolving models}.
\newblock {\em Phys. Rev. D}, 106(10):103511, 2022.

\bibitem{Hazra:2022rdl}
Dhiraj~Kumar Hazra, Akhil Antony, and Arman Shafieloo.
\newblock {One spectrum to cure them all: signature from early Universe solves
  major anomalies and tensions in cosmology}.
\newblock {\em JCAP}, 08(08):063, 2022.

\bibitem{Braglia:2021sun}
Matteo Braglia, Xingang Chen, and Dhiraj~Kumar Hazra.
\newblock {Uncovering the history of cosmic inflation from anomalies in cosmic
  microwave background spectra}.
\newblock {\em Eur. Phys. J. C}, 82(5):498, 2022.

\bibitem{Keeley:2020rmo}
Ryan~E. Keeley, Arman Shafieloo, Dhiraj~Kumar Hazra, and Tarun Souradeep.
\newblock {Inflation Wars: A New Hope}.
\newblock {\em JCAP}, 09:055, 2020.

\bibitem{Jiang:2023bsz}
Jun-Qian Jiang, Gen Ye, and Yun-Song Piao.
\newblock {Impact of the Hubble tension on the $r$-$n_s$ contour}.
\newblock 3 2023.

\bibitem{Martin:2013nzq}
J\'er\^ome Martin, Christophe Ringeval, Roberto Trotta, and Vincent Vennin.
\newblock {The Best Inflationary Models After Planck}.
\newblock {\em JCAP}, 03:039, 2014.

\bibitem{Lyth:1998xn}
David~H. Lyth and Antonio Riotto.
\newblock {Particle physics models of inflation and the cosmological density
  perturbation}.
\newblock {\em Phys. Rept.}, 314:1--146, 1999.

\bibitem{Linde:2007fr}
Andrei~D. Linde.
\newblock {Inflationary Cosmology}.
\newblock {\em Lect. Notes Phys.}, 738:1--54, 2008.

\bibitem{Baumann:2014nda}
Daniel Baumann and Liam McAllister.
\newblock {\em {Inflation and String Theory}}.
\newblock Cambridge Monographs on Mathematical Physics. Cambridge University
  Press, 5 2015.

\bibitem{Leach:2002ar}
Samuel~M. Leach, Andrew~R. Liddle, Jerome Martin, and Dominik~J Schwarz.
\newblock {Cosmological parameter estimation and the inflationary cosmology}.
\newblock {\em Phys. Rev. D}, 66:023515, 2002.

\bibitem{Boubekeur:2005zm}
Lotfi Boubekeur and David.~H. Lyth.
\newblock {Hilltop inflation}.
\newblock {\em JCAP}, 07:010, 2005.

\bibitem{Martin:2006rs}
Jerome Martin and Christophe Ringeval.
\newblock {Inflation after WMAP3: Confronting the Slow-Roll and Exact Power
  Spectra to CMB Data}.
\newblock {\em JCAP}, 08:009, 2006.

\bibitem{Moss:2007qd}
IanG. Moss and ChristopherM. Graham.
\newblock {Testing models of inflation with CMB non-gaussianity}.
\newblock {\em JCAP}, 11:004, 2007.

\bibitem{Bezrukov:2010jz}
F.~Bezrukov, A.~Magnin, M.~Shaposhnikov, and S.~Sibiryakov.
\newblock {Higgs inflation: consistency and generalisations}.
\newblock {\em JHEP}, 01:016, 2011.

\bibitem{Zhao:2011zb}
Wen Zhao and Qing-Guo Huang.
\newblock {Testing inflationary consistency relations by the potential CMB
  observations}.
\newblock {\em Class. Quant. Grav.}, 28:235003, 2011.

\bibitem{Martin:2014rqa}
Jerome Martin, Christophe Ringeval, and Vincent Vennin.
\newblock {How Well Can Future CMB Missions Constrain Cosmic Inflation?}
\newblock {\em JCAP}, 10:038, 2014.

\bibitem{Martin:2014lra}
Jerome Martin, Christophe Ringeval, Roberto Trotta, and Vincent Vennin.
\newblock {Compatibility of Planck and BICEP2 in the Light of Inflation}.
\newblock {\em Phys. Rev. D}, 90(6):063501, 2014.

\bibitem{Carrillo-Gonzalez:2014tia}
Mariana Carrillo-Gonz\'alez, Gabriel Germ\'an-Velarde, Alfredo Herrera-Aguilar,
  Juan~Carlos Hidalgo, and Roberto Sussman.
\newblock {Testing Hybrid Natural Inflation with BICEP2}.
\newblock {\em Phys. Lett. B}, 734:345--349, 2014.

\bibitem{Creminelli:2014oaa}
Paolo Creminelli, Diana L\'opez~Nacir, Marko Simonovi\'c, Gabriele Trevisan,
  and Matias Zaldarriaga.
\newblock {$\phi^2$ or Not $\phi^2$: Testing the Simplest Inflationary
  Potential}.
\newblock {\em Phys. Rev. Lett.}, 112(24):241303, 2014.

\bibitem{DiValentino:2016nni}
Eleonora Di~Valentino and Laura Mersini-Houghton.
\newblock {Testing Predictions of the Quantum Landscape Multiverse 1: The
  Starobinsky Inflationary Potential}.
\newblock {\em JCAP}, 03:002, 2017.

\bibitem{DiValentino:2016ziq}
Eleonora Di~Valentino and Laura Mersini-Houghton.
\newblock {Testing Predictions of the Quantum Landscape Multiverse 2: The
  Exponential Inflationary Potential}.
\newblock {\em JCAP}, 03:020, 2017.

\bibitem{Campista:2017ovq}
Marcela Campista, Micol Benetti, and Jailson Alcaniz.
\newblock {Testing non-minimally coupled inflation with CMB data: a Bayesian
  analysis}.
\newblock {\em JCAP}, 09:010, 2017.

\bibitem{Giare:2019snj}
William Giar\`e, Eleonora Di~Valentino, and Alessandro Melchiorri.
\newblock {Testing the inflationary slow-roll condition with tensor modes}.
\newblock {\em Phys. Rev. D}, 99(12):123522, 2019.

\bibitem{Dai:2019ejv}
Rui Dai and Yi~Zhu.
\newblock {Testing kinetically coupled inflation models with CMB distortions}.
\newblock {\em JCAP}, 05:017, 2020.

\bibitem{Baumann:2015xxa}
Daniel Baumann, Hayden Lee, and Guilherme~L. Pimentel.
\newblock {High-Scale Inflation and the Tensor Tilt}.
\newblock {\em JHEP}, 01:101, 2016.

\bibitem{Odintsov:2020ilr}
S.~D. Odintsov, V.~K. Oikonomou, and F.~P. Fronimos.
\newblock {Canonical scalar field inflation with string and $R^2$
  -corrections}.
\newblock {\em Annals Phys.}, 424:168359, 2021.

\bibitem{Giare:2020plo}
William Giar\`e, Fabrizio Renzi, and Alessandro Melchiorri.
\newblock {Higher-Curvature Corrections and Tensor Modes}.
\newblock {\em Phys. Rev. D}, 103(4):043515, 2021.

\bibitem{Oikonomou:2021kql}
V.~K. Oikonomou.
\newblock {A refined Einstein\textendash{}Gauss\textendash{}Bonnet inflationary
  theoretical framework}.
\newblock {\em Class. Quant. Grav.}, 38(19):195025, 2021.

\bibitem{Odintsov:2022cbm}
Sergei~D. Odintsov, Vasilis~K. Oikonomou, and Ratbay Myrzakulov.
\newblock {Spectrum of Primordial Gravitational Waves in Modified Gravities: A
  Short Overview}.
\newblock {\em Symmetry}, 14(4):729, 2022.

\bibitem{Namba:2015gja}
Ryo Namba, Marco Peloso, Maresuke Shiraishi, Lorenzo Sorbo, and Caner Unal.
\newblock {Scale-dependent gravitational waves from a rolling axion}.
\newblock {\em JCAP}, 01:041, 2016.

\bibitem{Peloso:2016gqs}
Marco Peloso, Lorenzo Sorbo, and Caner Unal.
\newblock {Rolling axions during inflation: perturbativity and signatures}.
\newblock {\em JCAP}, 09:001, 2016.

\bibitem{Pi:2019ihn}
Shi Pi, Misao Sasaki, and Ying-li Zhang.
\newblock {Primordial Tensor Perturbation in Double Inflationary Scenario with
  a Break}.
\newblock {\em JCAP}, 06:049, 2019.

\bibitem{Ozsoy:2020ccy}
Ogan \"Ozsoy.
\newblock {Synthetic Gravitational Waves from a Rolling Axion Monodromy}.
\newblock {\em JCAP}, 04:040, 2021.

\bibitem{Stewart:2007fu}
Andrew Stewart and Robert Brandenberger.
\newblock {Observational Constraints on Theories with a Blue Spectrum of Tensor
  Modes}.
\newblock {\em JCAP}, 08:012, 2008.

\bibitem{Mukohyama:2014gba}
Shinji Mukohyama, Ryo Namba, Marco Peloso, and Gary Shiu.
\newblock {Blue Tensor Spectrum from Particle Production during Inflation}.
\newblock {\em JCAP}, 08:036, 2014.

\bibitem{Giovannini:2015kfa}
Massimo Giovannini.
\newblock {The refractive index of relic gravitons}.
\newblock {\em Class. Quant. Grav.}, 33(12):125002, 2016.

\bibitem{Giovannini:2018dob}
Massimo Giovannini.
\newblock {Post-inflationary thermal histories and the refractive index of
  relic gravitons}.
\newblock {\em Phys. Rev. D}, 98(10):103509, 2018.

\bibitem{Giovannini:2018nkt}
Massimo Giovannini.
\newblock {Blue and violet graviton spectra from a dynamical refractive index}.
\newblock {\em Phys. Lett. B}, 789:502--507, 2019.

\bibitem{Giovannini:2018zbf}
Massimo Giovannini.
\newblock {The propagating speed of relic gravitational waves and their
  refractive index during inflation}.
\newblock {\em Eur. Phys. J. C}, 78(6):442, 2018.

\bibitem{Giare:2020vhn}
William Giar\`e and Alessandro Melchiorri.
\newblock {Probing the inflationary background of gravitational waves from
  large to small scales}.
\newblock {\em Phys. Lett. B}, 815:136137, 2021.

\bibitem{Giare:2020vss}
William Giar\`e and Fabrizio Renzi.
\newblock {Propagating speed of primordial gravitational waves}.
\newblock {\em Phys. Rev. D}, 102(8):083530, 2020.

\bibitem{Giare:2022wxq}
William Giar\`e, Matteo Forconi, Eleonora Di~Valentino, and Alessandro
  Melchiorri.
\newblock {Towards a reliable calculation of relic radiation from primordial
  gravitational waves}.
\newblock {\em Mon. Not. Roy. Astron. Soc.}, 520:2, 2023.

\bibitem{Baumgart:2021ptt}
Matthew Baumgart, Jonathan~J. Heckman, and Logan Thomas.
\newblock {CFTs blueshift tensor fluctuations universally}.
\newblock {\em JCAP}, 07(07):034, 2022.

\bibitem{Franciolini:2018ebs}
G.~Franciolini, G.~F. Giudice, D.~Racco, and A.~Riotto.
\newblock {Implications of the detection of primordial gravitational waves for
  the Standard Model}.
\newblock {\em JCAP}, 05:022, 2019.

\bibitem{DEramo:2019tit}
Francesco D'Eramo and Kai Schmitz.
\newblock {Imprint of a scalar era on the primordial spectrum of gravitational
  waves}.
\newblock {\em Phys. Rev. Research.}, 1:013010, 2019.

\bibitem{Caldwell:2018giq}
Robert~R. Caldwell, Tristan~L. Smith, and Devin G.~E. Walker.
\newblock {Using a Primordial Gravitational Wave Background to Illuminate New
  Physics}.
\newblock {\em Phys. Rev. D}, 100(4):043513, 2019.

\bibitem{Clarke:2020bil}
Thomas~J. Clarke, Edmund~J. Copeland, and Adam Moss.
\newblock {Constraints on primordial gravitational waves from the Cosmic
  Microwave Background}.
\newblock {\em JCAP}, 10:002, 2020.

\bibitem{Caprini_2018}
Chiara Caprini and Daniel~G Figueroa.
\newblock Cosmological backgrounds of gravitational waves.
\newblock {\em CQG}, 35(16):163001, jul 2018.

\bibitem{Allen:1997ad}
Bruce Allen and Joseph~D. Romano.
\newblock {Detecting a stochastic background of gravitational radiation: Signal
  processing strategies and sensitivities}.
\newblock {\em Phys. Rev. D}, 59:102001, 1999.

\bibitem{Smith:2006nka}
Tristan~L. Smith, Elena Pierpaoli, and Marc Kamionkowski.
\newblock {A new cosmic microwave background constraint to primordial
  gravitational waves}.
\newblock {\em Phys. Rev. Lett.}, 97:021301, 2006.

\bibitem{Boyle:2007zx}
Latham~A. Boyle and Alessandra Buonanno.
\newblock {Relating gravitational wave constraints from primordial
  nucleosynthesis, pulsar timing, laser interferometers, and the CMB:
  Implications for the early Universe}.
\newblock {\em Phys. Rev. D}, 78:043531, 2008.

\bibitem{Kuroyanagi:2014nba}
Sachiko Kuroyanagi, Tomo Takahashi, and Shuichiro Yokoyama.
\newblock {Blue-tilted Tensor Spectrum and Thermal History of the Universe}.
\newblock {\em JCAP}, 02:003, 2015.

\bibitem{Ben-Dayan:2019gll}
Ido Ben-Dayan, Brian Keating, David Leon, and Ira Wolfson.
\newblock {Constraints on scalar and tensor spectra from $N_{eff}$}.
\newblock {\em JCAP}, 06:007, 2019.

\bibitem{Aich:2019obd}
Moumita Aich, Yin-Zhe Ma, Wei-Ming Dai, and Jun-Qing Xia.
\newblock {How much primordial tensor mode is allowed?}
\newblock {\em Phys. Rev. D}, 101(6):063536, 2020.

\bibitem{Cabass:2015jwe}
Giovanni Cabass, Luca Pagano, Laura Salvati, Martina Gerbino, Elena Giusarma,
  and Alessandro Melchiorri.
\newblock {Updated Constraints and Forecasts on Primordial Tensor Modes}.
\newblock {\em Phys. Rev. D}, 93(6):063508, 2016.

\bibitem{Vagnozzi:2020gtf}
Sunny Vagnozzi.
\newblock {Implications of the NANOGrav results for inflation}.
\newblock {\em Mon. Not. Roy. Astron. Soc.}, 502(1):L11--L15, 2021.

\bibitem{Benetti:2021uea}
Micol Benetti, Leila~Lobato Graef, and Sunny Vagnozzi.
\newblock {Primordial gravitational waves from NANOGrav: A broken power-law
  approach}.
\newblock {\em Phys. Rev. D}, 105(4):043520, 2022.

\bibitem{Calcagni:2020tvw}
Gianluca Calcagni and Sachiko Kuroyanagi.
\newblock {Stochastic gravitational-wave background in quantum gravity}.
\newblock {\em JCAP}, 03:019, 2021.

\bibitem{Oikonomou:2022ijs}
V.~K. Oikonomou.
\newblock {Amplification of the Primordial Gravitational Waves Energy Spectrum
  by a Kinetic Scalar in $F(R)$ Gravity}.
\newblock {\em Astropart. Phys.}, 144:102777, 2023.

\bibitem{Barrow:1993ad}
John~D. Barrow, J.~P. Mimoso, and M.~R. de~Garcia~Maia.
\newblock {Amplification of gravitational waves in scalar - tensor theories of
  gravity}.
\newblock {\em Phys. Rev. D}, 48:3630, 1993.
\newblock [Erratum: Phys.Rev.D 51, 5967 (1995)].

\bibitem{Peng:2021zon}
Zhi-Zhang Peng, Chengjie Fu, Jing Liu, Zong-Kuan Guo, and Rong-Gen Cai.
\newblock {Gravitational waves from resonant amplification of curvature
  perturbations during inflation}.
\newblock {\em JCAP}, 10:050, 2021.

\bibitem{Ota:2022hvh}
Atsuhisa Ota, Misao Sasaki, and Yi~Wang.
\newblock {Scale-invariant enhancement of gravitational waves during
  inflation}.
\newblock 9 2022.

\bibitem{Odintsov:2022sdk}
S.~D. Odintsov and V.~K. Oikonomou.
\newblock {Amplification of Primordial Gravitational Waves by a Geometrically
  Driven non-canonical Reheating Era}.
\newblock {\em Fortsch. Phys.}, 70(5):2100167, 2022.

\bibitem{Capurri:2020qgz}
Giulia Capurri, Nicola Bartolo, Davide Maino, and Sabino Matarrese.
\newblock {Let Effective Field Theory of Inflation flow: stochastic generation
  of models with red/blue tensor tilt}.
\newblock {\em JCAP}, 11:037, 2020.

\bibitem{Canas-Herrera:2021sjs}
Guadalupe Ca\~nas Herrera and Fabrizio Renzi.
\newblock {Current and future constraints on single-field
  \ensuremath{\alpha}-attractor models}.
\newblock {\em Phys. Rev. D}, 104(10):103512, 2021.

\bibitem{Odintsov:2023aaw}
S.~D. Odintsov, V.~K. Oikonomou, and F.~P. Fronimos.
\newblock {Inflationary Dynamics and Swampland Criteria for Modified
  Gauss-Bonnet Gravity Compatible with GW170817}.
\newblock 3 2023.

\bibitem{Oikonomou:2023bah}
V.~K. Oikonomou.
\newblock {Effects of the axion through the Higgs portal on primordial
  gravitational waves during the electroweak breaking}.
\newblock {\em Phys. Rev. D}, 107(6):064071, 2023.

\bibitem{Fronimos:2023tim}
F.~P. Fronimos and S.~A. Venikoudis.
\newblock {Inflationary phenomenology of non-minimally coupled
  Einstein-Chern-Simons gravity}.
\newblock 2 2023.

\bibitem{Cai:2022lec}
Yong Cai.
\newblock {Generating enhanced parity-violating gravitational waves during
  inflation with violation of the null energy condition}.
\newblock {\em Phys. Rev. D}, 107(6):063512, 2023.

\bibitem{Oikonomou:2022irx}
V.~K. Oikonomou.
\newblock {Effects of a pre-inflationary de Sitter bounce on the primordial
  gravitational waves in f(R) gravity theories}.
\newblock {\em Nucl. Phys. B}, 984:115985, 2022.

\bibitem{Gangopadhyay:2022vgh}
Mayukh~R. Gangopadhyay, Hussain~Ahmed Khan, and Yogesh.
\newblock {A case study of small field inflationary dynamics in the
  Einstein\textendash{}Gauss\textendash{}Bonnet framework in the light of
  GW170817}.
\newblock {\em Phys. Dark Univ.}, 40:101177, 2023.

\bibitem{Odintsov:2022hxu}
S.~D. Odintsov and V.~K. Oikonomou.
\newblock {Chirality of gravitational waves in Chern-Simons f(R) gravity
  cosmology}.
\newblock {\em Phys. Rev. D}, 105(10):104054, 2022.

\bibitem{Odintsov:2020mkz}
S.~D. Odintsov, V.~K. Oikonomou, F.~P. Fronimos, and S.~A. Venikoudis.
\newblock {GW170817-compatible constant-roll
  Einstein\textendash{}Gauss\textendash{}Bonnet inflation and
  non-Gaussianities}.
\newblock {\em Phys. Dark Univ.}, 30:100718, 2020.

\bibitem{Galloni:2022mok}
Giacomo Galloni, Nicola Bartolo, Sabino Matarrese, Marina Migliaccio, Angelo
  Ricciardone, and Nicola Vittorio.
\newblock {Updated constraints on amplitude and tilt of the tensor primordial
  spectrum}.
\newblock {\em JCAP}, 04:062, 2023.

\bibitem{DeAngelis:2021afq}
Mariaveronica De~Angelis, Laria Figurato, and Giovanni Montani.
\newblock {Quantum dynamics of the isotropic universe in metric f(R) gravity}.
\newblock {\em Phys. Rev. D}, 104(2):024054, 2021.

\bibitem{Kallosh:2022ggf}
Renata Kallosh and Andrei Linde.
\newblock {Hybrid cosmological attractors}.
\newblock {\em Phys. Rev. D}, 106(2):023522, 2022.

\bibitem{Braglia:2022phb}
Matteo Braglia, Andrei Linde, Renata Kallosh, and Fabio Finelli.
\newblock {Hybrid \ensuremath{\alpha}-attractors, primordial black holes and
  gravitational wave backgrounds}.
\newblock {\em JCAP}, 04:033, 2023.

\bibitem{Vagnozzi:2022qmc}
Sunny Vagnozzi and Abraham Loeb.
\newblock {The Challenge of Ruling Out Inflation via the Primordial Graviton
  Background}.
\newblock {\em Astrophys. J. Lett.}, 939(2):L22, 2022.

\bibitem{Senatore:2010wk}
Leonardo Senatore and Matias Zaldarriaga.
\newblock {The Effective Field Theory of Multifield Inflation}.
\newblock {\em JHEP}, 04:024, 2012.

\bibitem{Starobinsky:2001xq}
Alexei~A. Starobinsky, Shinji Tsujikawa, and Jun'ichi Yokoyama.
\newblock {Cosmological perturbations from multifield inflation in generalized
  Einstein theories}.
\newblock {\em Nucl. Phys. B}, 610:383--410, 2001.

\bibitem{Tsujikawa:2002qx}
Shinji Tsujikawa, David Parkinson, and Bruce~A. Bassett.
\newblock {Correlation - consistency cartography of the double inflation
  landscape}.
\newblock {\em Phys. Rev. D}, 67:083516, 2003.

\bibitem{DiMarco:2002eb}
Fabrizio Di~Marco, Fabio Finelli, and Robert Brandenberger.
\newblock {Adiabatic and isocurvature perturbations for multifield generalized
  Einstein models}.
\newblock {\em Phys. Rev. D}, 67:063512, 2003.

\bibitem{Kaiser:2010ps}
David~I. Kaiser.
\newblock {Conformal Transformations with Multiple Scalar Fields}.
\newblock {\em Phys. Rev. D}, 81:084044, 2010.

\bibitem{Achucarro:2010da}
Ana Achucarro, Jinn-Ouk Gong, Sjoerd Hardeman, Gonzalo~A. Palma, and Subodh~P.
  Patil.
\newblock {Features of heavy physics in the CMB power spectrum}.
\newblock {\em JCAP}, 01:030, 2011.

\bibitem{vandeBruck:2010yw}
Carsten van~de Bruck, David~F. Mota, and Joel~M. Weller.
\newblock {Embedding DBI inflation in scalar-tensor theory}.
\newblock {\em JCAP}, 03:034, 2011.

\bibitem{Kaiser:2013sna}
David~I. Kaiser and Evangelos~I. Sfakianakis.
\newblock {Multifield Inflation after Planck: The Case for Nonminimal
  Couplings}.
\newblock {\em Phys. Rev. Lett.}, 112(1):011302, 2014.

\bibitem{vandeBruck:2015tna}
Carsten van~de Bruck, Tomi Koivisto, and Chris Longden.
\newblock {Disformally coupled inflation}.
\newblock {\em JCAP}, 03:006, 2016.

\bibitem{vandeBruck:2015xpa}
Carsten van~de Bruck and Laura~Elena Paduraru.
\newblock {Simplest extension of Starobinsky inflation}.
\newblock {\em Phys. Rev. D}, 92:083513, 2015.

\bibitem{vandeBruck:2016vlw}
Carsten van~de Bruck, Tomi Koivisto, and Chris Longden.
\newblock {Non-Gaussianity in multi-sound-speed disformally coupled inflation}.
\newblock {\em JCAP}, 02:029, 2017.

\bibitem{Carrilho:2018ffi}
Pedro Carrilho, David Mulryne, John Ronayne, and Tommi Tenkanen.
\newblock {Attractor Behaviour in Multifield Inflation}.
\newblock {\em JCAP}, 06:032, 2018.

\bibitem{Achucarro:2019pux}
Ana Ach\'ucarro, Edmund~J. Copeland, Oksana Iarygina, Gonzalo~A. Palma,
  Dong-Gang Wang, and Yvette Welling.
\newblock {Shift-symmetric orbital inflation: Single field or multifield?}
\newblock {\em Phys. Rev. D}, 102(2):021302, 2020.

\bibitem{Pinol:2020kvw}
Lucas Pinol.
\newblock {Multifield inflation beyond $N_\mathrm{field}=2$: non-Gaussianities
  and single-field effective theory}.
\newblock {\em JCAP}, 04:002, 2021.

\bibitem{Achucarro:2018vey}
Ana Ach\'ucarro and Gonzalo~A. Palma.
\newblock {The string swampland constraints require multi-field inflation}.
\newblock {\em JCAP}, 02:041, 2019.

\bibitem{vandeBruck:2021xkm}
Carsten van~de Bruck and Richard Daniel.
\newblock {Inflation and scale-invariant R$^2$ gravity}.
\newblock {\em Phys. Rev. D}, 103(12):123506, 2021.

\bibitem{DeAngelis:2023fdu}
Mariaveronica De~Angelis and Carsten van~de Bruck.
\newblock {Adiabatic and isocurvature perturbations in extended theories with
  non--minimally coupled fields}.
\newblock 4 2023.

\bibitem{Weinberg:2004kf}
Steven Weinberg.
\newblock {Must cosmological perturbations remain non-adiabatic after
  multi-field inflation?}
\newblock {\em Phys. Rev. D}, 70:083522, 2004.

\bibitem{Tsujikawa:2000wc}
Shinji Tsujikawa and Hiroki Yajima.
\newblock {New constraints on multifield inflation with nonminimal coupling}.
\newblock {\em Phys. Rev. D}, 62:123512, 2000.

\bibitem{Kaiser:2010yu}
David~I. Kaiser and Audrey~T. Todhunter.
\newblock {Primordial Perturbations from Multifield Inflation with Nonminimal
  Couplings}.
\newblock {\em Phys. Rev. D}, 81:124037, 2010.

\bibitem{Frazer:2013zoa}
Jonathan Frazer.
\newblock {Predictions in multifield models of inflation}.
\newblock {\em JCAP}, 01:028, 2014.

\bibitem{Achucarro:2012fd}
Ana Ach\'ucarro, Jinn-Ouk Gong, Gonzalo~A. Palma, and Subodh~P. Patil.
\newblock {Correlating features in the primordial spectra}.
\newblock {\em Phys. Rev. D}, 87(12):121301, 2013.

\bibitem{vandeBruck:2014ata}
Carsten van~de Bruck and Mathew Robinson.
\newblock {Power Spectra beyond the Slow Roll Approximation in Theories with
  Non-Canonical Kinetic Terms}.
\newblock {\em JCAP}, 08:024, 2014.

\bibitem{Dias:2015rca}
Mafalda Dias, Jonathan Frazer, and David Seery.
\newblock {Computing observables in curved multifield models of
  inflation\textemdash{}A guide (with code) to the transport method}.
\newblock {\em JCAP}, 12:030, 2015.

\bibitem{Dias:2016rjq}
Mafalda Dias, Jonathan Frazer, David~J. Mulryne, and David Seery.
\newblock {Numerical evaluation of the bispectrum in multiple field
  inflation\textemdash{}the transport approach with code}.
\newblock {\em JCAP}, 12:033, 2016.

\bibitem{Braglia:2020fms}
Matteo Braglia, Dhiraj~Kumar Hazra, L.~Sriramkumar, and Fabio Finelli.
\newblock {Generating primordial features at large scales in two field models
  of inflation}.
\newblock {\em JCAP}, 08:025, 2020.

\bibitem{Braglia:2021ckn}
Matteo Braglia, Xingang Chen, and Dhiraj~Kumar Hazra.
\newblock {Comparing multi-field primordial feature models with the Planck
  data}.
\newblock {\em JCAP}, 06:005, 2021.

\bibitem{Cabass:2022ymb}
Giovanni Cabass, Mikhail~M. Ivanov, Oliver H.~E. Philcox, Marko Simonovi\'c,
  and Matias Zaldarriaga.
\newblock {Constraints on multifield inflation from the BOSS galaxy survey}.
\newblock {\em Phys. Rev. D}, 106(4):043506, 2022.

\bibitem{Geller:2022nkr}
Sarah~R. Geller, Wenzer Qin, Evan McDonough, and David~I. Kaiser.
\newblock {Primordial black holes from multifield inflation with nonminimal
  couplings}.
\newblock {\em Phys. Rev. D}, 106(6):063535, 2022.

\bibitem{Wang:2022eop}
Dong-Gang Wang, Guilherme~L. Pimentel, and Ana Ach\'ucarro.
\newblock {Bootstrapping Multi-Field Inflation: non-Gaussianities from light
  scalars revisited}.
\newblock 12 2022.

\bibitem{Iacconi:2023slv}
Laura Iacconi and David~J. Mulryne.
\newblock {Multi-field inflation with large scalar fluctuations:
  non-Gaussianity and perturbativity}.
\newblock 4 2023.

\bibitem{Qin:2023lgo}
Wenzer Qin, Sarah~R. Geller, Shyam Balaji, Evan McDonough, and David~I. Kaiser.
\newblock {Planck Constraints and Gravitational Wave Forecasts for Primordial
  Black Hole Dark Matter Seeded by Multifield Inflation}.
\newblock 3 2023.

\bibitem{Freytsis:2022aho}
Marat Freytsis, Soubhik Kumar, Grant~N. Remmen, and Nicholas~L. Rodd.
\newblock {Multifield Positivity Bounds for Inflation}.
\newblock 10 2022.

\bibitem{Cicoli:2021yhb}
Michele Cicoli, Veronica Guidetti, Francesco Muia, Francisco~G. Pedro, and
  Gian~Paolo Vacca.
\newblock {On the choice of entropy variables in multifield inflation}.
\newblock {\em Class. Quant. Grav.}, 40(2):025008, 2023.

\bibitem{Guerrero:2020lng}
Merce Guerrero, Diego Rubiera-Garcia, and Diego Saez-Chillon~Gomez.
\newblock {Constant roll inflation in multifield models}.
\newblock {\em Phys. Rev. D}, 102:123528, 2020.

\bibitem{Garcia-Saenz:2019njm}
Sebastian Garcia-Saenz, Lucas Pinol, and S\'ebastien Renaux-Petel.
\newblock {Revisiting non-Gaussianity in multifield inflation with curved field
  space}.
\newblock {\em JHEP}, 01:073, 2020.

\bibitem{Nguyen:2019kbm}
Rachel Nguyen, Jorinde van~de Vis, Evangelos~I. Sfakianakis, John~T. Giblin,
  and David~I. Kaiser.
\newblock {Nonlinear Dynamics of Preheating after Multifield Inflation with
  Nonminimal Couplings}.
\newblock {\em Phys. Rev. Lett.}, 123(17):171301, 2019.

\bibitem{Li:2019zbk}
Xi-Bin Li, Xiao-Gang Zheng, and Jian-Yang Zhu.
\newblock {Spectra and entropy of multifield warm inflation}.
\newblock {\em Phys. Rev. D}, 99(4):043528, 2019.

\bibitem{Bernardeau:2002jy}
Francis Bernardeau and Jean-Philippe Uzan.
\newblock {NonGaussianity in multifield inflation}.
\newblock {\em Phys. Rev. D}, 66:103506, 2002.

\bibitem{Kaiser:2012ak}
David~I. Kaiser, Edward~A. Mazenc, and Evangelos~I. Sfakianakis.
\newblock {Primordial Bispectrum from Multifield Inflation with Nonminimal
  Couplings}.
\newblock {\em Phys. Rev. D}, 87:064004, 2013.

\bibitem{McAllister:2012am}
Liam McAllister, Sebastien Renaux-Petel, and Gang Xu.
\newblock {A Statistical Approach to Multifield Inflation: Many-field
  Perturbations Beyond Slow Roll}.
\newblock {\em JCAP}, 10:046, 2012.

\bibitem{Peterson:2011yt}
Courtney~M. Peterson and Max Tegmark.
\newblock {Testing multifield inflation: A geometric approach}.
\newblock {\em Phys. Rev. D}, 87(10):103507, 2013.

\bibitem{Dias:2012nf}
Mafalda Dias, Jonathan Frazer, and Andrew~R. Liddle.
\newblock {Multifield consequences for D-brane inflation}.
\newblock {\em JCAP}, 06:020, 2012.
\newblock [Erratum: JCAP 03, E01 (2013)].

\bibitem{Kehagias:2012td}
A.~Kehagias and A.~Riotto.
\newblock {The Four-point Correlator in Multifield Inflation, the Operator
  Product Expansion and the Symmetries of de Sitter}.
\newblock {\em Nucl. Phys. B}, 868:577--595, 2013.

\bibitem{Leung:2012ve}
Godfrey Leung, Ewan R.~M. Tarrant, Christian~T. Byrnes, and Edmund~J. Copeland.
\newblock {Reheating, Multifield Inflation and the Fate of the Primordial
  Observables}.
\newblock {\em JCAP}, 09:008, 2012.

\bibitem{Meyers:2010rg}
Joel Meyers and Navin Sivanandam.
\newblock {Non-Gaussianities in Multifield Inflation: Superhorizon Evolution,
  Adiabaticity, and the Fate of fnl}.
\newblock {\em Phys. Rev. D}, 83:103517, 2011.

\bibitem{Price:2014ufa}
Layne~C. Price, Hiranya~V. Peiris, Jonathan Frazer, and Richard Easther.
\newblock {Gravitational wave consistency relations for multifield inflation}.
\newblock {\em Phys. Rev. Lett.}, 114(3):031301, 2015.

\bibitem{Battefeld:2011yj}
Diana Battefeld, Thorsten Battefeld, Christian Byrnes, and David Langlois.
\newblock {Beauty is Distractive: Particle production during multifield
  inflation}.
\newblock {\em JCAP}, 08:025, 2011.

\bibitem{Kaiser:2015usz}
David~I. Kaiser.
\newblock {Nonminimal Couplings in the Early Universe: Multifield Models of
  Inflation and the Latest Observations}.
\newblock {\em Fundam. Theor. Phys.}, 183:41--57, 2016.

\bibitem{Ashcroft:2002ap}
P.~R. Ashcroft, C.~van~de Bruck, and A.~C. Davis.
\newblock {Suppression of entropy perturbations in multi-field inflation on the
  brane}.
\newblock {\em Phys. Rev. D}, 66:121302, 2002.

\bibitem{Paliathanasis:2021fxi}
Andronikos Paliathanasis and Genly Leon.
\newblock {Global dynamics of the hyperbolic Chiral-Phantom model}.
\newblock {\em Eur. Phys. J. Plus}, 137(1):165, 2022.

\bibitem{Paliathanasis:2020abu}
Andronikos Paliathanasis and Genly Leon.
\newblock {Asymptotic behavior of $N$-fields Chiral Cosmology}.
\newblock {\em Eur. Phys. J. C}, 80(9):847, 2020.

\bibitem{Paliathanasis:2020wjl}
Andronikos Paliathanasis.
\newblock {Dynamics of Chiral Cosmology}.
\newblock {\em Class. Quant. Grav.}, 37(19):195014, 2020.

\bibitem{Christodoulidis:2021vye}
Perseas Christodoulidis and Andronikos Paliathanasis.
\newblock {$\mathcal{N}$-field cosmology in hyperbolic field space: stability
  and general solutions}.
\newblock {\em JCAP}, 05:038, 2021.

\bibitem{Piao:2006nm}
Yun-Song Piao.
\newblock {On perturbation spectra of N-flation}.
\newblock {\em Phys. Rev. D}, 74:047302, 2006.

\bibitem{Rinaldi:2023mdf}
Massimiliano Rinaldi, Chiara Cecchini, Anish Ghoshal, and Debangshu Mukherjee.
\newblock {Scale-invariant inflation}.
\newblock {\em J. Phys. Conf. Ser.}, 2531(1):012012, 2023.

\bibitem{Ijaz:2023cvc}
Nadir Ijaz, Maria Mehmood, and Mansoor~Ur Rehman.
\newblock {The Stochastic Gravitational-Wave Background from Primordial Black
  Holes in R-Symmetric $SU(5)$ Inflation}.
\newblock 8 2023.

\bibitem{Planck:2018vyg}
N.~Aghanim et~al.
\newblock {Planck 2018 results. VI. Cosmological parameters}.
\newblock {\em Astron. Astrophys.}, 641:A6, 2020.
\newblock [Erratum: Astron.Astrophys. 652, C4 (2021)].

\bibitem{Planck:2018nkj}
N.~Aghanim et~al.
\newblock {Planck 2018 results. I. Overview and the cosmological legacy of
  Planck}.
\newblock {\em Astron. Astrophys.}, 641:A1, 2020.

\bibitem{ACT:2020gnv}
Simone Aiola et~al.
\newblock {The Atacama Cosmology Telescope: DR4 Maps and Cosmological
  Parameters}.
\newblock {\em JCAP}, 12:047, 2020.

\bibitem{ACT:2020frw}
Steve~K. Choi et~al.
\newblock {The Atacama Cosmology Telescope: a measurement of the Cosmic
  Microwave Background power spectra at 98 and 150 GHz}.
\newblock {\em JCAP}, 12:045, 2020.

\bibitem{SPT-3G:2022hvq}
L.~Balkenhol et~al.
\newblock {A Measurement of the CMB Temperature Power Spectrum and Constraints
  on Cosmology from the SPT-3G 2018 TT/TE/EE Data Set}.
\newblock 12 2022.

\bibitem{SPT-3G:2014dbx}
B.~A. Benson et~al.
\newblock {SPT-3G: A Next-Generation Cosmic Microwave Background Polarization
  Experiment on the South Pole Telescope}.
\newblock {\em Proc. SPIE Int. Soc. Opt. Eng.}, 9153:91531P, 2014.

\bibitem{SPT-3G:2021eoc}
D.~Dutcher et~al.
\newblock {Measurements of the E-mode polarization and temperature-E-mode
  correlation of the CMB from SPT-3G 2018 data}.
\newblock {\em Phys. Rev. D}, 104(2):022003, 2021.

\bibitem{Easther:2013rva}
Richard Easther, Jonathan Frazer, Hiranya~V. Peiris, and Layne~C. Price.
\newblock {Simple predictions from multifield inflationary models}.
\newblock {\em Phys. Rev. Lett.}, 112:161302, 2014.

\bibitem{Price:2014xpa}
Layne~C. Price, Jonathan Frazer, Jiajun Xu, Hiranya~V. Peiris, and Richard
  Easther.
\newblock {MultiModeCode: An efficient numerical solver for multifield
  inflation}.
\newblock {\em JCAP}, 03:005, 2015.

\bibitem{Lewis:1999bs}
Antony Lewis, Anthony Challinor, and Anthony Lasenby.
\newblock {Efficient computation of CMB anisotropies in closed FRW models}.
\newblock {\em Astrophys. J.}, 538:473--476, 2000.

\bibitem{Howlett:2012mh}
Cullan Howlett, Antony Lewis, Alex Hall, and Anthony Challinor.
\newblock {CMB power spectrum parameter degeneracies in the era of precision
  cosmology}.
\newblock {\em JCAP}, 04:027, 2012.

\bibitem{Blas:2011rf}
Diego Blas, Julien Lesgourgues, and Thomas Tram.
\newblock {The Cosmic Linear Anisotropy Solving System (CLASS) II:
  Approximation schemes}.
\newblock {\em JCAP}, 07:034, 2011.

\bibitem{Gordon:2000hv}
Christopher Gordon, David Wands, Bruce~A. Bassett, and Roy Maartens.
\newblock {Adiabatic and entropy perturbations from inflation}.
\newblock {\em Phys. Rev. D}, 63:023506, 2000.

\bibitem{PhysRevD.67.063512}
F.~Di~Marco, F.~Finelli, and R.~Brandenberger.
\newblock Adiabatic and isocurvature perturbations for multifield generalized
  einstein models.
\newblock {\em Phys. Rev. D}, 67:063512, Mar 2003.

\bibitem{Langlois:2008mn}
David Langlois and Sebastien Renaux-Petel.
\newblock {Perturbations in generalized multi-field inflation}.
\newblock {\em JCAP}, 04:017, 2008.

\bibitem{Langlois_2008}
David Langlois and Sébastien Renaux-Petel.
\newblock Perturbations in generalized multi-field inflation.
\newblock {\em Journal of Cosmology and Astroparticle Physics}, 2008(04):017,
  apr 2008.

\bibitem{bib:wands-2002}
David Wands, Nicola Bartolo, Sabino Matarrese, and Antonio Riotto.
\newblock {An Observational test of two-field inflation}.
\newblock {\em Phys. Rev. D}, 66:043520, 2002.

\bibitem{Silk:1986vc}
Joseph Silk and Michael~S. Turner.
\newblock {Double Inflation}.
\newblock {\em Phys. Rev. D}, 35:419, 1987.

\bibitem{Polarski:1992dq}
David Polarski and Alexei~A. Starobinsky.
\newblock {Spectra of perturbations produced by double inflation with an
  intermediate matter dominated stage}.
\newblock {\em Nucl. Phys. B}, 385:623--650, 1992.

\bibitem{Roberts:1994ap}
David Roberts, Andrew~R. Liddle, and David~H. Lyth.
\newblock {False vacuum inflation with a quartic potential}.
\newblock {\em Phys. Rev. D}, 51:4122--4128, 1995.

\bibitem{Langlois:1999dw}
David Langlois.
\newblock {Correlated adiabatic and isocurvature perturbations from double
  inflation}.
\newblock {\em Phys. Rev. D}, 59:123512, 1999.

\bibitem{Leach:2000yw}
Samuel~M. Leach and Andrew~R. Liddle.
\newblock {Inflationary perturbations near horizon crossing}.
\newblock {\em Phys. Rev. D}, 63:043508, 2001.

\bibitem{Leach:2001zf}
Samuel~M Leach, Misao Sasaki, David Wands, and Andrew~R Liddle.
\newblock {Enhancement of superhorizon scale inflationary curvature
  perturbations}.
\newblock {\em Phys. Rev. D}, 64:023512, 2001.

\bibitem{Jain:2007au}
Rajeev~Kumar Jain, Pravabati Chingangbam, and L.~Sriramkumar.
\newblock {On the evolution of tachyonic perturbations at super-Hubble scales}.
\newblock {\em JCAP}, 10:003, 2007.

\bibitem{Jain:2008dw}
Rajeev~Kumar Jain, Pravabati Chingangbam, Jinn-Ouk Gong, L.~Sriramkumar, and
  Tarun Souradeep.
\newblock {Punctuated inflation and the low CMB multipoles}.
\newblock {\em JCAP}, 01:009, 2009.

\bibitem{Jain:2009pm}
Rajeev~Kumar Jain, Pravabati Chingangbam, L.~Sriramkumar, and Tarun Souradeep.
\newblock {The tensor-to-scalar ratio in punctuated inflation}.
\newblock {\em Phys. Rev. D}, 82:023509, 2010.

\bibitem{Kallosh:2014rga}
Renata Kallosh, Andrei Linde, and Diederik Roest.
\newblock {Large field inflation and double $\alpha$-attractors}.
\newblock {\em JHEP}, 08:052, 2014.

\bibitem{Ragavendra:2020old}
H.~V. Ragavendra, Debika Chowdhury, and L.~Sriramkumar.
\newblock {Suppression of scalar power on large scales and associated
  bispectra}.
\newblock {\em Phys. Rev. D}, 106(4):043535, 2022.

\bibitem{Aghanim:2019ame}
N.~Aghanim et~al.
\newblock {Planck 2018 results. V. CMB power spectra and likelihoods}.
\newblock {\em Astron. Astrophys.}, 641:A5, 2020.

\bibitem{Aghanim:2018eyx}
N.~Aghanim et~al.
\newblock {Planck 2018 results. VI. Cosmological parameters}.
\newblock {\em Astron. Astrophys.}, 641:A6, 2020.

\bibitem{Akrami:2018vks}
Y.~Akrami et~al.
\newblock {Planck 2018 results. I. Overview and the cosmological legacy of
  Planck}.
\newblock 2018.

\bibitem{Aghanim:2018oex}
N.~Aghanim et~al.
\newblock {Planck 2018 results. VIII. Gravitational lensing}.
\newblock 7 2018.

\bibitem{bib:bardeen-1980}
James~M. Bardeen.
\newblock Gauge-invariant cosmological perturbations.
\newblock {\em Phys. Rev. D}, 22:1882--1905, Oct 1980.

\bibitem{vandeBruck:2016rfv}
Carsten van~de Bruck and Chris Longden.
\newblock {Running of the Running and Entropy Perturbations During Inflation}.
\newblock {\em Phys. Rev. D}, 94(2):021301, 2016.

\bibitem{Chen:2012ja}
Xingang Chen and Christophe Ringeval.
\newblock {Searching for Standard Clocks in the Primordial Universe}.
\newblock {\em JCAP}, 08:014, 2012.

\bibitem{Chen:2014joa}
Xingang Chen and Mohammad~Hossein Namjoo.
\newblock {Standard Clock in Primordial Density Perturbations and Cosmic
  Microwave Background}.
\newblock {\em Phys. Lett. B}, 739:285--292, 2014.

\bibitem{Chen:2014cwa}
Xingang Chen, Mohammad~Hossein Namjoo, and Yi~Wang.
\newblock {Models of the Primordial Standard Clock}.
\newblock {\em JCAP}, 02:027, 2015.

\bibitem{Chen:2015lza}
Xingang Chen, Mohammad~Hossein Namjoo, and Yi~Wang.
\newblock {Quantum Primordial Standard Clocks}.
\newblock {\em JCAP}, 02:013, 2016.

\bibitem{Peterson:2010np}
Courtney~M. Peterson and Max Tegmark.
\newblock {Testing Two-Field Inflation}.
\newblock {\em Phys. Rev. D}, 83:023522, 2011.

\bibitem{Torrado:2020xyz}
Jesus Torrado and Antony Lewis.
\newblock {Cobaya: Code for Bayesian Analysis of hierarchical physical models}.
\newblock {\em arXiv:2005.05290}, 5 2020.

\bibitem{Lewis:2002ah}
Antony Lewis and Sarah Bridle.
\newblock {Cosmological parameters from CMB and other data: A Monte Carlo
  approach}.
\newblock {\em Phys. Rev. D}, 66:103511, 2002.

\bibitem{Neal:2005}
R.~M. {Neal}.
\newblock {Taking Bigger Metropolis Steps by Dragging Fast Variables}.
\newblock {\em ArXiv Mathematics e-prints}, February 2005.

\end{thebibliography}
%\end{thebibliography}

% Please avoid comments such as "For a review'', "For some examples",
% "and references therein" or move them in the text. In general,
% please leave only references in the bibliography and move all
% accessory text in footnotes.

% Also, please have only one work for each \bibitem.

\end{document}